\documentclass[10pt,twocolumn]{IEEEtran}

\usepackage{adjustbox}

\usepackage{diagbox}
\usepackage{enumitem}
\usepackage{amsmath,graphics,amssymb,epsfig,subfigure,color,cite}
\usepackage{array}
\usepackage{multirow}
\usepackage{enumerate}
\usepackage{algorithmic}
\usepackage{algorithm}
\usepackage{bm}
\usepackage{float}
\usepackage{makecell}
\usepackage{lipsum}
\usepackage{capt-of}
\usepackage{amsmath}
\usepackage{amssymb}

\newtheorem{thm}{Theorem}

\floatname{algorithm}{Algorithm}

\begin{document}
        \title{Importance-Aware Semantic Communication in MIMO-OFDM Systems Using Vision Transformer}
	 \author{Joohyuk Park, Yongjeong Oh, Jihun Park, and Yo-Seb Jeon, \IEEEmembership{Member,~IEEE}
    \thanks{Joohyuk Park, Yongjeong Oh, Jihun Park, and Yo-Seb Jeon are with the Department of Electrical Engineering, POSTECH, Pohang, Gyeongbuk 37673, South Korea (e-mail: joohyuk.park@postech.ac.kr; yongjeongoh@postech.ac.kr; jihun.park@postech.ac.kr; yoseb.jeon@postech.ac.kr).}
    }
	\vspace{-2mm}
	
	\maketitle
 
	\vspace{-12mm}

    \begin{abstract} 
        This paper presents a novel importance-aware quantization, subcarrier mapping, and power allocation (IA-QSMPA) framework for semantic communication in multiple-input multiple-output orthogonal frequency division multiplexing (MIMO-OFDM) systems, empowered by a pretrained Vision Transformer (ViT). The proposed framework exploits attention-based importance extracted from a pretrained ViT to jointly optimize quantization levels, subcarrier mapping, and power allocation. Specifically, IA-QSMPA maps semantically important features to high-quality subchannels and allocates resources in accordance with their contribution to task performance and communication latency. To efficiently solve the resulting nonconvex optimization problem, a block coordinate descent algorithm is employed. The framework is further extended to operate under finite blocklength transmission, where communication errors may occur. In this setting, a segment-wise linear approximation of the channel dispersion penalty is introduced to enable efficient joint optimization under practical constraints. Simulation results on a multi-view image classification task using the MVP-N dataset demonstrate that IA-QSMPA significantly outperforms conventional methods in both ideal and finite blocklength transmission scenarios, achieving superior task performance and communication efficiency.
    \end{abstract}

    
    \section{Introduction}\label{Sec:Intro}
        Recent advancements in semantic communication have redirected the focus of communication systems from accurate bit reconstruction to the effective delivery of task-relevant information \cite{SC_1}. 
        Unlike traditional systems that prioritize minimizing bit-level errors, semantic communication systems seek to maximize task performance by preserving the intended meaning of transmitted content. 
        This paradigm shift is particularly beneficial in resource-constrained communication scenarios, 
        where efficient use of frequency and time resources is essential for ensuring reliable and timely communication \cite{Application_2, Application_3}.
        To support this goal, most semantic communication systems adopt a joint source-channel coding (JSCC) strategy, in which task-relevant semantic features are directly transmitted over wireless channels using deep neural networks. 
        In this approach, the source and channel encoding/decoding processes are combined into a single neural network model trained under wireless channel conditions such as additive white Gaussian noise (AWGN) and Rayleigh fading.
        This approach has shown strong performance across diverse tasks, including image transmission \cite{image_trans_2, image_trans_3}, text transmission \cite{DeepSC, ReAllo-T}, and speech transmission \cite{DeepSC-S}.  
        However, JSCC typically relies on analog transmission, which limits compatibility with existing digital communication infrastructure. Moreover, analog approaches are generally more sensitive to noise and offer less scalability and flexibility than their digital counterparts.        

        Digital semantic communication has received increasing attention as a more flexible and standard-compatible approach to realizing semantic communication. These systems aim to represent semantic features using finite-valued representations that can be readily mapped to digital symbols, enabling reliable and efficient transmission using conventional digital communication transceivers. 
        For example, in \cite{NECST, DSC_Fixed_bit_6, EC_2, DSC_Fixed_bit_1, DSC_Fixed_bit_7}, the source data was first compressed into semantic features using a JSCC encoder, and these features were then quantized and converted into bit sequences. 
        In particular, in \cite{DSC_Fixed_bit_1}, the communication environment was modeled using binary symmetric channels (BSCs), where random bit flip probabilities were sampled and used during training to improve system robustness. Building on this, in \cite{DSC_Fixed_bit_7}, the bit flip probability was further treated as a learnable parameter, allowing the system to estimate the error sensitivity of each feature individually.
        In contrast, in \cite{DSC_Symbol_1}, semantic features were directly mapped to modulated symbols without being converted into bits. Meanwhile, in \cite{DSC_Fixed_bit_5, FATD}, the non-differentiability of certain modules during training was addressed by employing continuous relaxation techniques. Specifically, in \cite{DSC_Fixed_bit_5}, a differentiable quantizer was proposed, whereas in \cite{FATD}, differentiable modulation and demodulation processes were developed.
        Unfortunately, these approaches are limited in that they rely on simplified communication assumptions, such as AWGN channel and single-input single-output system.

        To overcome these limitations, semantic communication tailored for more practical system configurations has been studied in \cite{Digital_JSCC_MIMO_3,Digital_JSCC_MIMO_5,Digital_JSCC_OFDM}.        
        For example,  in \cite{Digital_JSCC_MIMO_3},  semantic communication in multiple-input multiple-output (MIMO) systems was investigated for image transmission. Similarly, in \cite{Digital_JSCC_MIMO_5}, semantic communication in orthogonal frequency division multiplexing (OFDM) system was investigated based on a digital JSCC approach.
        A channel-adaptive digital JSCC approach for OFDM systems was developed in \cite{Digital_JSCC_OFDM}, where BSCs were utilized to equivalently represent the joint effect of OFDM modulation and demodulation.
        A common feature of these methods is their dependence on end-to-end (E2E) training. Although E2E learning has proven effective in controlled environments, it lacks robustness when training and testing environments are mismatched. Moreover, E2E training becomes impractical in complex settings such as multi-modal, multi-task, and collaborative inference \cite{VQA,MDCEI,MVID}, as it requires retraining for every possible scenario or environment change.
        These limitations hinder the scalability and practical applicability of E2E semantic communication approaches in real-world systems such as cellular and IoT sensor networks.

        Semantic communication approaches that do not rely on E2E training have been explored in \cite{TF_IA_1, TF_IA_2, TF_IA_3, TF_IA_4, TF_IA_5}, primarily focusing on integrating semantic importance into transceiver designs. 
        In \cite{TF_IA_1}, a bit-level interleaving strategy was introduced to assign more critical bits to more robust positions within the modulation constellation, thereby improving resilience to channel impairments. In \cite{TF_IA_2, TF_IA_3}, important features were transmitted when the signal-to-noise ratio (SNR) was high, thus preserving essential information. In \cite{TF_IA_4}, a pretrained vision transformer (ViT) encoder was employed to identify task-relevant image patches, and the transmission of each quantized patch was selectively determined based on its semantic significance without relying on E2E training. Building on this approach, in \cite{TF_IA_5}, a quantization strategy was proposed in which different quantization levels were assigned to patches according to their task relevance.
        Although these prior works have successfully incorporated semantic importance into feature selection and quantization strategies, optimal semantic feature transmission within practical communication systems remains largely unexplored. In particular, importance-aware optimization methods for quantization-level adjustment, subcarrier mapping, and power control in realistic communication scenarios, such as MIMO-OFDM systems, represent significant and open research challenges.

        To bridge the gap toward realizing training-free and practical semantic communication, we propose a novel framework for importance-aware semantic communication in MIMO-OFDM systems using ViT. The proposed framework is termed importance-aware quantization, subcarrier mapping, and power allocation (IA-QSMPA). This framework jointly performs quantization, subcarrier mapping, and power control by leveraging semantic importance extracted from the attention scores of a pretrained ViT. By assigning communication resources in proportion to the importance of each semantic feature, IA-QSMPA simultaneously enhances task accuracy and reduces transmission latency, thereby improving both the reliability and efficiency of semantic communication.
        The main contributions of this paper are summarized as follows:

    \begin{figure*}[t]
        \centering 
        {\epsfig{file=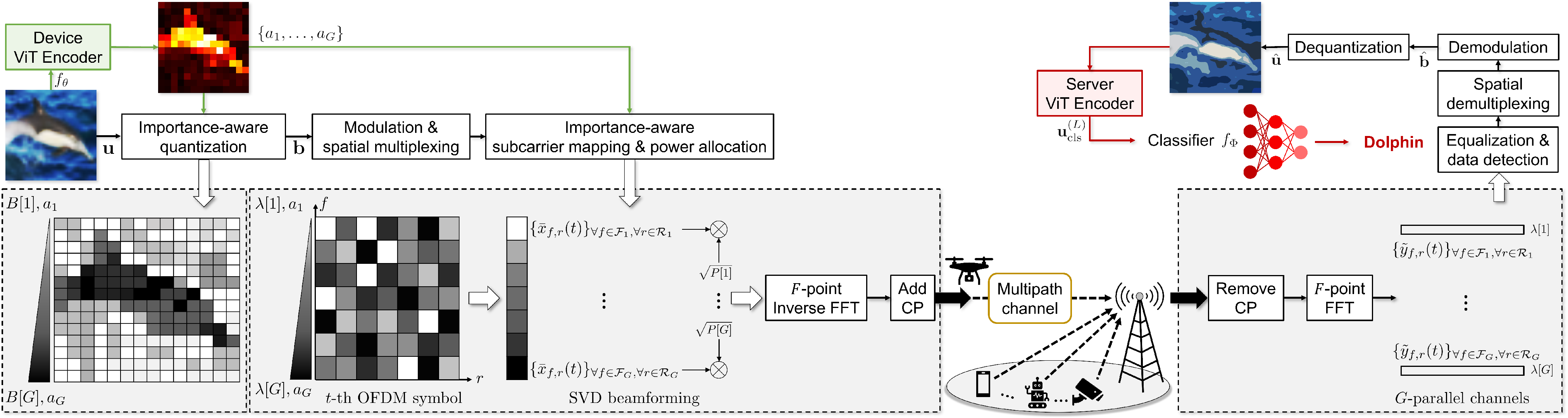, width=18cm}} 
        \caption{Illustration of the proposed IA-QSMPA framework for ViT-based semantic communication in MIMO-OFDM systems.}
        \label{fig:System}
    \end{figure*}  

    \begin{itemize}
    \item 
    We investigate a joint optimization problem for ViT-based semantic communication in MIMO-OFDM systems, which integrates subcarrier mapping with quantization-bit allocation and power control. To the best of our knowledge, this is the first study addressing the joint optimization of quantization bits, symbol mapping, and power allocation. This work extends previous research by combining importance-aware subcarrier mapping (IASM) \cite{TF_IA_2, TF_IA_3} and importance-aware quantization (IAQ) \cite{TF_IA_5} into a unified problem suitable for practical MIMO-OFDM systems.

    \item We propose a novel IA-QSMPA framework for ViT-based semantic communication in MIMO-OFDM systems under an ideal transmission scenario. The core idea is to quantify the importance of each semantic feature (e.g., image patch) using the mean attention score extracted from a pretrained ViT. Semantic features with higher importance scores are then mapped to subchannels exhibiting stronger channel gains to enhance transmission reliability. Subsequently, based on this importance-driven mapping, quantization bits and power levels for each semantic feature are jointly optimized to minimize the weighted quantization error and communication latency, where the weights are designed as monotonically increasing functions of the mean attention scores.  To efficiently solve the resulting nonconvex optimization problem, we adopt a block coordinate descent (BCD) algorithm \cite{BCD}, yielding a near-optimal solution.

    \item We extend our IA-QSMPA framework to operate under the finite blocklength transmission scenario, which inherently involves a nonzero probability of communication errors. In this setting, the framework is adapted to reduce communication latency while maintaining task performance in the presence of transmission uncertainty. To enable efficient optimization, we approximate the channel dispersion term appearing in the finite blocklength achievable rate expression using a segment-wise linear function. This approximation transforms the joint bit and power optimization problem into a tractable form that can be efficiently solved via a BCD algorithm.

    \item Using simulations, we demonstrate the superiority of our IA-QSMPA framework over existing transmission methods for multi-view image classification using the MVP-N \cite{MVP_N} dataset. The results show that the proposed framework provides significant gains in the considered task compared to the existing methods. These results highlight the effectiveness of integrating semantic importance into transmission optimization in MIMO-OFDM semantic communication systems.

 
    \end{itemize}

    \section{System Model}\label{Sec:Model}
    Consider a point-to-point MIMO-OFDM semantic communication system for wireless image transmission, where a device transmits an image to a server that performs a dedicated machine learning task (e.g., image classification), as illustrated in Fig.~\ref{fig:System}. Given input data ${\bf{u}} \in \mathbb{R}^{H\times W\times C}$, where $H$, $W$, and $C$ represent the height, width, and number of channels, respectively, we first process the image using a ViT encoder $f_{\theta}$ deployed at the device, parameterized by weights $\theta$. Following the approach in \cite{TF_IA_5}, this encoder extracts patch-wise attention scores, which serve as semantic importance indicators. These scores guide the IAQ process, wherein more bits are allocated to semantically critical regions to preserve task-relevant information. This results in a variable-length bit sequence that encodes the image content with spatially adaptive precision. Further implementation details are provided in Sec.~\ref{Sec:IA_QSMPA_error_free} and Sec.~\ref{Sec:Modified_IA_QSMPA_error_robust}.

    Based on the quantization results, the total number of generated bits is given by $D\sum_{i=1}^{G}B[i]$, where $D=P^2C$, $(P, P)$ denotes the patch size, $G = \frac{HW}{P^2}$ is the number of patches, and $B[i]$ indicates the number of quantization bits assigned to the $i$-th patch.
    Through this process, the $i$-th patch is converted into $D B[i]$ information bits, denoted by ${\bf b}\in\{0,1\}^{D \sum_{i=1}^{G} B[i]}$.
    After channel encoding and symbol modulation, these bits are transformed into a symbol sequence of length $L[i]$, represented as
    \begin{align}\label{eq:per_block_symbol_1}
        {\bf x}^{i}= [x_1^{i}, \dots, x_{L[i]}^{i}]^{\rm T}
        \in \mathbb{C}^{L[i]},~\forall i\in \{1,\ldots,G\},
    \end{align}
    where $x^{i}_l$ is the $l$-th entry of ${\bf x}^{i}$.
    For clarity, we define the $i$-th block as the transmission unit corresponding to the $i$-th patch and refer to the modulated sequence ${\bf x}^{i}$ as belonging to this block.

    The overall frequency-domain transmitted signal, consisting of all $G$ blocks, is then given by
    \begin{align}\label{eq:per_block_symbol_2}
        {\bf x} = \left[[{\bf x}^{1}]^{\rm T},\ldots, [{\bf x}^{G}]^{\rm T} \right]^{\rm T} \in \mathbb{C}^{L_{\rm tot}},
    \end{align}
    where $L_{\rm tot}=\sum_{i=1}^{G}L[i]$. 
    If spatial multiplexing with $N_s$ streams is applied to ${\bf x}$, the signal is first reorganized into a two-dimensional (2D) matrix, denoted by ${\sf Reshape} \left({\bf x}\right) \in \mathbb{C}^{N_s \times \frac{L_{\rm tot}}{N_s}}$. Subsequently, IASM, which will be described in detail in Sec.~\ref{Sec:IA_QSMPA_error_free}-A, is applied. Accordingly, the frequency-domain transmitted signal at subcarrier $f$ of the $t$-th OFDM symbol in \eqref{eq:per_block_symbol_2} is expressed as
    \begin{align}\label{eq:per_block_symbol_3}
        \bar{\bf x}_f(t) = \left[\bar{x}_{f,1}(t),\ldots,\bar{x}_{f,N_s}(t)\right]^{\rm T} \in \mathbb{C}^{N_s},
    \end{align}
    for all $f \in \{1,\ldots,F\}$, where $F$ denote the total number of subcarriers. 
    Accordingly, after applying importance-aware power allocation, the frequency-domain transmitted signal can be modified as
    \begin{align}\label{eq:per_block_symbol_4}
        \tilde{\bf x}_f(t) =  {\bf W}_{{\rm T}, f}(t){\bf P}_f^{1/2}(t)\bar{\bf x}_f(t) \in \mathbb{C}^{N_{\rm tx}}, 
    \end{align}
    where $N_{\rm tx}$ is the number of transmit antenna, ${\bf P}^{1/2}_f(t) = {\sf diag}(P^{1/2}_{f,1}(t),\ldots, P^{1/2}_{f,N_s}(t)) \in \mathbb{R}^{N_s \times N_s}$ is the power allocation matrix, and ${\bf W}_{{\rm T}, f}(t) \in \mathbb{C}^{N_{\rm tx} \times N_s}$ is the transmit beamforming matrix. This signal is converted into the time domain by  applying an $F$-point inverse fast Fourier transform (FFT) and then transmitted after adding the cyclic prefix (CP).

    
    
    At the server, the CP is first removed from the received time-domain signal. Subsequently, an $F$-point FFT is applied to convert the signal into the frequency domain. Let ${\bf y}_{f}(t) \in \mathbb{C}^{N_{\rm rx}}$ denote the received signal at subcarrier $f$ of the $t$-th OFDM symbol, where $N_{\rm rx}$ is the number of receive antennas. This signal is given by
    \begin{align}\label{eq:per_block_symbol_7}
        {\bf y}_{f}(t) &= {\bf H}_{f}(t)\tilde{\bf x}_f(t) +{\bf v}_{f}(t),
    \end{align}
    where ${\bf H}_{f}(t) \in \mathbb{C}^{N_{\rm rx} \times N_{\rm tx}}$ is the frequency-domain MIMO channel matrix at subcarrier $f$ of the $t$-th OFDM symbol. Each entry of ${\bf H}_{f}(t)$ is assumed to follow the distribution $[{\bf H}_{f}(t)]_{r_1, r_2} \sim \mathcal{CN}(0, \sigma_H^2)$, where $\sigma_H^2$ denotes the channel variance. The noise vector ${\bf v}_{f}(t) \sim \mathcal{CN}({\bf 0}_{N_{\rm rx}}, \sigma^2 {\bf I}_{N_{\rm rx}})$ represents additive white Gaussian noise (AWGN) in the frequency domain with variance $\sigma^2$ per dimension. 

    To analyze the optimal operation of the MIMO-OFDM semantic communication systems, both the device and server are assumed to have perfect knowledge of the frequency-domain channel state information and the noise variance $\sigma^2$. Moreover, the channel is assumed to remain constant over the coherence time, i.e.,  $\mathbf{H}_{f}(t) = \mathbf{H}_{f}, \forall t \in \{1,\ldots,T\}$.
    Based on this, singular value decomposition (SVD) beamforming \cite{Block_fading} decomposes the channel into orthogonal parallel subchannels, enabling interference-free transmission for each data stream. As a result, at each $t$-th OFDM symbol, a total of $N_sF$ parallel channels are formed, and the received signal after receive beamforming can be obtained as
    \begin{align}\label{eq:per_block_symbol_8}
        \tilde{\bf y}_{f}(t) &= {\bf W}_{{\rm R}, f}^{\rm H}{\bf y}_{f}(t) \notag \\
        &\overset{(a)}{=} {\boldsymbol \Lambda}_{f}{\bf P}^{1/2}_{f}(t)\bar{\bf x}_{f}(t)+\tilde{\bf v}_{f}(t) \in \mathbb{C}^{N_{s}},
    \end{align}
    where $\tilde{\bf v}_f(t) = {\bf W}_{{\rm R}, f}^{\rm H}{\bf v}_f(t)$ denotes an effective noise vector, ${\bf W}_{{\rm R}, f} \in \mathbb{C}^{N_{\rm rx} \times N_s}$ denotes the receive beamforming matrix, and ${\bf \Lambda}_f = {\sf diag}(\lambda_{f,1},\ldots, \lambda_{f,N_s}) \in \mathbb{R}^{N_s \times N_s}$ represents the singular value matrix, whose diagonal elements correspond to the channel gains at subcarrier $f$. Note that $(a)$ follows from the fact that $\mathbf{W}_{{\rm T}, f}(t)=\mathbf{W}_{{\rm T}, f}$ is obtained by selecting the first $N_s$ columns of the right unitary matrix from the SVD of $\mathbf{H}_f$, while $\mathbf{W}_{{\rm R}, f}$ is formed by taking the first $N_s$ columns of the left unitary matrix.

    Then, the element-wise received signal associated with the $i$-th block (i.e., the $i$-th patch) is expressed as
    \begin{align}\label{eq:per_block_symbol_9}
        \tilde{y}_{f,r}(t) = \lambda_{f,r}\sqrt{P_{f,r}(t)}\bar{x}_{f,r}(t)+\tilde{v}_{f,r}(t),
    \end{align}
    for all $t \in \mathcal{T}_i$, $f \in \mathcal{F}_i$, and $r \in \mathcal{R}_i$.
    Here, $\mathcal{T}_i$, $\mathcal{F}_i$, and $\mathcal{R}_i$ denote the respective sets of OFDM symbols, subcarriers, and antenna domain resources assigned to $\mathbf{x}^i$. The total number of allocated subchannels in the $i$-th block satisfies $|\mathcal{T}_i||\mathcal{F}_i||\mathcal{R}_i| = L[i]$. To ensure that the channel remains constant during transmission, it is assumed that the number of OFDM symbols assigned to each block is smaller than the coherence time (i.e., $|\mathcal{T}_i| < T$, $\forall i$). Furthermore, we assume that the number of frequency-antenna subchannels allocated to each block at every OFDM symbol is uniformly distributed (i.e., $|\mathcal{F}_i||\mathcal{R}_i| = \frac{N_sF}{G}$, $\forall i$).

    The received signal $\tilde{y}_{f,r}(t)$ in \eqref{eq:per_block_symbol_9} is equalized by dividing by the effective channel gain $\lambda_{f,r}\sqrt{P_{f,r}(t)}$. The equalized signals are then demultiplexed to reconstruct the transmitted symbols $\bf x$. These symbols are demodulated and decoded to recover the encoded bitstream $\hat{\bf b}$, which is subsequently dequantized based on the bit allocation $\{B[i]\}_{i=1}^G$ to obtain the reconstructed image $\hat{\bf u}$. Finally, $\hat{\bf u}$ is passed through the pretrained ViT encoder to obtain the class token ${\bf u}^{(L)}_{\rm cls}$. This token is then fed into the classifier $f_\Phi$, parameterized by weights $\Phi$, to perform the downstream task (e.g., image classification task). 

    \section{Proposed IA-QSMPA Framework under Ideal Transmission Scenario}\label{Sec:IA_QSMPA_error_free}
    In this section, we present a novel IA-QSMPA framework to enable importance-aware image transmission in MIMO-OFDM systems. The focus is specifically on optimizing a transmission strategy under an ideal transmission scenario. An extension of our framework for a more practical scenario will be discussed in Sec.~\ref{Sec:Modified_IA_QSMPA_error_robust}.

    \subsection{Importance-Aware Subcarrier Mapping (IASM)}\label{CGOSM}
    In MIMO-OFDM semantic communication systems, the joint optimization of subcarrier mapping, quantization bit allocation, and power control poses significant complexity due to the coupling among these variables. To simplify the process, we first assign subcarriers to semantic symbols in an importance-aware manner, prioritizing more important symbols. Based on this, joint bit and power allocation is then performed to efficiently assign resources according to symbol importance.

    As discussed in Sec.~\ref{Sec:Model}, the channel gains $\left\{\lambda_{f,r}\right\}_{\forall f, \forall r}$ are derived through SVD beamforming. In semantic communications, a key step for enhancing task-related performance is to map the modulated symbol sequence $\mathbf{x}^i$ to appropriate transmission channels. To guide this process, the mean attention score $a_i$ of the $i$-th patch is computed using a pretrained ViT, serving as a measure of the semantic importance of the corresponding symbols. 
    Patches with larger $a_i$ values, indicating higher importance, are transmitted through channels with higher gains. This allocation strategy prioritizes the reliable delivery of semantically critical information.

    To facilitate this mapping, the $N_sF$ distinct channel gains are first sorted in ascending order based on their magnitudes. These ordered gains are then partitioned into $G$ blocks. As a result, the subchannels in each block exhibit similar channel gains, allowing them to be approximated by their average value. This block-wise organization enables patch-level abstraction of the channel, which significantly reduces implementation complexity by allowing power allocation to be performed at the patch level instead of the symbol level. A uniform power level is thus assigned to all symbols within a given block. Meanwhile, the mapping ensures that patches with higher semantic importance are transmitted over blocks with stronger average channel gains, thereby enhancing the overall transmission reliability. Mathematically, at the $t$-th OFDM symbol, this mapping is characterized by
    \begin{align}
         \lambda[i] &= \frac{\sum_{f \in \mathcal{F}_i}^{}\sum_{r \in \mathcal{R}_i}^{}\lambda_{f,r}}{N_sF/G}, \label{eq:per_block_symbol_10_1}\\
        \lambda[i] &\leq \lambda[j],~\forall i \leq j, \label{eq:per_block_symbol_10_1_}\\
        P[i] &= P_{f,r}(t), ~\text{if}~ f\in \mathcal{F}_i, r \in \mathcal{R}_i, \label{eq:per_block_symbol_10_2}
    \end{align}
    where $\lambda[i]$ denotes the average channel gain of the subchannels allocated to the $i$-th block, and serves as the equivalent channel gain representing the quality of the channel associated with that block. Based on the IASM, the resulting equivalent channel gains $\{\lambda[i]\}_{i=1}^{G}$ are used to determine the optimal power allocation $\{P[i]\}_{i=1}^{G}$ and bit allocation $\{B[i]\}_{i=1}^{G}$ for each block, ensuring that resource allocation aligns with the semantic importance of the transmitted symbols.

    \subsection{Patch-Wise Quantization and Power Allocation}\label{Sec:IA_QSMPA}
    {In traditional MIMO-OFDM systems, transmitting an image in a capacity-achieving manner typically involves distributing data across parallel subchannels without considering the semantic structure of the content. In contrast, semantic communication prioritizes the preservation of task-relevant meaning, which motivates a more structured approach to data handling.
    In this context, block-wise data processing remains central, and associating each block with a semantic unit, such as an image patch, offers significant benefits. This design allows the system to leverage variations in equivalent channel gains by assigning more informative patches to stronger subchannels and less critical ones to weaker channels. Such importance-aware allocation improves task performance and also enables practical functionalities, including selective retransmission and low-latency delivery of essential content. These capabilities are especially useful in real-time applications with stringent delay requirements.}
    
    Building upon this motivation, we consider a MIMO-OFDM semantic communication system where the joint optimization of semantic task performance and communication latency is essential. In the proposed system, the $i$-th block contains $DB[i]$ bits of information, which are transmitted over a dedicated parallel channel with equivalent channel gain $\lambda[i]$ and allocated power $P[i]$. 
    Based on this configuration, the worst-case communication latency can be expressed as
    \begin{align}\label{eq:latency}
    E_{\rm L} = \underset{i}{\rm max}~\frac{DB[i]}{\Delta f \log_2(1+\gamma[i])},
    \end{align}
    where $\gamma[i] = \frac{P[i] \lambda^2[i]}{\sigma^2}$ denotes the received SNR of the $i$-th block, $\Delta f = \frac{N_sF}{G}\Delta f_0$ represents the effective bandwidth assigned per block, and $\Delta f_0$ denotes the subcarrier spacing.    
    This formulation highlights the role of $E_{\rm L}$ as the transmission bottleneck, since a larger value of $E_{\rm L}$ implies a greater number of OFDM symbols $|\mathcal{T}_i|$ required to complete the transmission of the most time-consuming block. Minimizing this worst-case latency is therefore essential to ensure timely delivery of all semantically meaningful information.


    Under an ideal transmission scenario, semantic distortion primarily results from quantization. To effectively minimize this distortion, we adopt a weighted quantization error model that accounts for the varying semantic importance of different patches. Building on our prior work \cite{TF_IA_5}, the quantization error under a uniform quantizer is upper-bounded by
    \begin{align}\label{eq:grouping}
    E_{\rm Q} = \sum_{i=1}^{G} I[i] \frac{D(u_{\rm max} - u_{\rm min})^2}{4}  4^{-B[i]},
    \end{align}
    where $u_{\rm min}$ and $u_{\rm max}$ denote the minimum and maximum values of the elements in $\mathbf{u}$, respectively. The weight term $I[i]$ is defined as a monotonically increasing function of $a_i$, and thus reflects its semantic significance.
    A higher value of $I[i]$ indicates that distortion in the $i$-th block has a more pronounced impact on task performance and consequently necessitates finer quantization. As a result, a larger bit allocation $B[i]$ is required, which increases the associated transmission delay. 
    Therefore, the weight $I[i]$ serves to control how strongly the allocation strategy prioritizes reducing quantization distortion over minimizing communication latency.
    Building upon this trade-off, the joint optimization of quantization accuracy and communication latency is formulated as
    \begin{subequations}
    \begin{align}
    ({\bf P}_1)~~&\underset{\{B[i], P[i]\}^{G}_{i=1}}{\rm min}~  E_{\rm L}+{E}_{\rm Q}, \label{eq:P_1}\\
    {\rm s.t.}~~~ &\sum_{i=1}^{G}DB[i] = B_{\rm target}, \label{P_1_C_1}\\
    &\sum_{i=1}^{G}\frac{N_sF}{G}P[i] = P_{\rm tot}, \label{P_1_C_2}\\
    &B[i] \in \left\{B_{\rm min},\ldots, B_{\rm max}\right\}, \forall i \in \left\{1,\ldots, G\right\}, \label{P_1_C_3}\\
    &P[i] \geq 0, \forall i, \label{P_1_C_4}
    \end{align}
    \end{subequations}
    where $B_{\rm min}$ and $B_{\rm max}$ denote the minimum and maximum number of quantization bits assignable to each patch, respectively. Based on these bounds, $B_{\rm target}$ represents the total number of bits allocated for quantizing all patches within a single image, satisfying $B_{\rm min}GD \leq B_{\rm target} \leq B_{\rm max}GD$. In addition, $P_{\rm tot}$ denotes the total transmit power available per OFDM symbol across all blocks. The additional bit overhead required to transmit the quantizer configuration $\{B[i]\}_{i=1}^{G}$ and the values of $u_{\rm min}$ and $u_{\rm max}$ is negligible relative to the total bit budget $B_{\rm target}$ and is thus omitted from the formulation.

    In the problem $({\bf P}_1)$, the variables $\{B[i]\}_{i=1}^G$ are discrete, while $\{P[i]\}_{i=1}^G$ are continuous, resulting in a mixed-integer nonlinear programming (MINLP) problem.
    To facilitate a more tractable solution, we first relax the discrete bit variables ${B[i]}$ to take continuous values. Additionally, we reformulate the maximum term in $E_{\rm L}$ by introducing a auxiliary variable $y$, which allows us to handle the latency component in a more analytically manageable form, i.e.,
    \begin{subequations}
    \begin{align}
        ({\bf P}_2)~~&\underset{\{B[i], P[i]\}^{G}_{i=1}, y}{\rm min}~ y+{E}_{\rm Q}, \label{eq:P_2}\\
        {\rm s.t.}~~~ &\frac{DB[i]}{\Delta f \log_2(1+\gamma[i])} \leq y, \forall i \in \left\{1,\ldots, G\right\}, \label{P_2_C_1} \\
        &B_{\rm min} \leq B[i] \leq B_{\rm max}, \label{P_2_C_2}\\
        &\eqref{P_1_C_1}, \eqref{P_1_C_2}. \notag
    \end{align}
    \end{subequations}
    In this problem, the constraint in \eqref{P_1_C_4} is omitted, as it is implicitly satisfied by \eqref{P_2_C_1} under the practical assumption that $y > 0$ and $B_{\rm min} \geq 1$.

    Although the problem $({\bf P}_2)$ remains nonconvex due to variable coupling and nonlinear constraints, its structure enables efficient optimization using the BCD algorithm \cite{BCD}. The key advantage of the BCD algorithm is that it decomposes the original problem into smaller subproblems, each optimizing a single group of variables while keeping the others fixed. Specifically, the optimization problem exhibits a group-wise convex structure:
    \begin{itemize}
    \item When $\left\{P[i]\right\}_{i=1}^G$ and $y$ are fixed, the problem reduces to a convex optimization with respect to $\left\{B[i]\right\}_{i=1}^G$.
    \item When $\left\{B[i]\right\}_{i=1}^G$ are fixed, the problem becomes jointly convex in $y$ and $\left\{P[i]\right\}_{i=1}^G$.
    \end{itemize}
    Each subproblem is convex and therefore an optimal solution can be obtained by applying the Karush-Kuhn-Tucker (KKT) conditions. 
    Furthermore, the objective function in \eqref{eq:P_2} is continuously differentiable and strictly convex with respect to each group of optimization variables. The constraints \eqref{P_1_C_1}, \eqref{P_1_C_2}, \eqref{P_2_C_1}, and \eqref{P_2_C_2} are compact and convex with respect to their corresponding optimization variables, which ensures that the BCD algorithm converges to a near-optimal solution of the problem $({\bf P}_2)$. The solution for each variable in iteration $k$ of the BCD algorithm is provided in the following theorem:

    \vspace{1mm}
    \begin{thm}
        The optimal solution of the problem $({\bf P}_2)$ in iteration $k$ of the BCD algorithm is
    \begin{align}
        B^{(k)}[i] &= {\rm min}\left\{\bar{B}^{(k)}_{\rm max}[i], {\rm max}\left\{B_{\rm min}, \hat{B}^{(k)}[i] \right\}\!\right\}, \label{eq:opt_sol_IA_QSMPA_1_1}
    \end{align}
    where
    \begin{align}
        \bar{B}^{(k)}_{\rm max}[i] &= \underset{}{\rm min} \left(\frac{y^{(k-1)}\Delta {f}}{D}\log_2\left(1+\gamma^{(k-1)}[i]\right), B_{\rm max}\right) ,\label{eq:opt_sol_IA_QSMPA_2_1}\\
        \hat{B}^{(k)}[i] &= \frac{1}{2}\log_2\left(\frac{ I[i]  \ln 2(u_{{\rm max}}-u_{{\rm min}})^2}{2\nu^{\star(k)}}\right),\label{eq:opt_sol_IA_QSMPA_2_2}\\
        \gamma^{(k)}[i] &= \frac{P^{(k)}[i]\lambda^2[i]}{\sigma^2}, \label{eq:opt_sol_IA_QSMPA_2_3}
    \end{align}
    for all $i \in \{1,\ldots, G\}$, $k \in \{1,\ldots, K\}$. The optimal Lagrange multiplier $\nu^{\star (k)}$ is determined to satisfy the following equality:
    \begin{align}\label{eq:La_mul_1}
        \sum_{i=1}^{G}B^{(k)}[i] = \frac{B_{\rm target}}{D}.
    \end{align} 
    In addition, the optimal auxiliary variable $y^{(k)}$ is obtained by solving the nonlinear equation:
    \begin{align}
        \sum_{i=1}^{G}\frac{\sigma^2}{\lambda^2[i]}\left(2^{\frac{DB^{(k)}[i]}{y^{(k)} \Delta f}}-1\right) &= \frac{P_{\rm tot}G}{N_sF}. \label{eq:dual_eq_2} 
    \end{align}
    Based on this result, the optimal power allocation can be determined as
    \begin{align}
        P^{(k)}[i] &= \frac{\sigma^2}{\lambda^2[i]}\left(2^{\frac{DB^{(k)}[i]}{y^{(k)}\Delta {f}}}-1\right), \label{eq:opt_sol_IA_QSMPA_1_3}
    \end{align}
    for all $i \in \{1,\ldots, G\}$, $k \in \{1,\ldots, K\}$.
    Once $\{B^{(k)}[i]\}_{i=1}^{G}$ and $y^{(k)}$ are determined, the associated Lagrange multiplier $\tau^{\star (k)}$ can be computed as
    \begin{align}
        \tau^{\star (k)} &= \left(\sum_{i=1}^{G}\frac{ \sigma^2 D B^{(k)}[i]\ln 2}{\Delta f_0\lambda^2[i]\left\{y^{(k)}\right\}^2} \cdot 2^{\frac{D B^{(k)}[i]}{y^{(k)} \Delta f}}\right)^{-1}. \label{eq:dual_eq_1} 
    \end{align}
    \end{thm}
    \begin{IEEEproof}
         See Appendix A.
    \end{IEEEproof}
    \vspace{1mm}

    The optimal Lagrange multiplier $\nu^{\star(k)}$, along with the auxiliary variable $y^{(k)}$, can be efficiently determined using numerical methods such as the fast water-filling (WF) algorithm~\cite{Water_filling_3} or the bisection search algorithm~\cite{Water_filling_2}. 
    Once these values are obtained, the optimal bit allocation $B^{(k)}[i]$ are computed via \eqref{eq:opt_sol_IA_QSMPA_1_1}, and the corresponding power allocation $P^{(k)}[i]$ is updated using \eqref{eq:opt_sol_IA_QSMPA_1_3}.

    Theorem 1 shows that when $\bar{B}^{(k)}_{\rm max}[i] = B_{\rm max}$ for all $i$ in \eqref{eq:opt_sol_IA_QSMPA_2_1}, the optimal quantization level is a monotonically increasing function of the weight $I[i]$. Consequently, the allocation in \eqref{eq:opt_sol_IA_QSMPA_1_1} assigns higher quantization levels to patches with greater mean attention scores, consistent with our previous work \cite{TF_IA_5}.
    However, as indicated in \eqref{eq:opt_sol_IA_QSMPA_2_1}, the value of $\bar{B}^{(k)}_{\rm max}[i]$ is determined by the previously updated auxiliary variable $y^{(k-1)}$. To support the latency constraint associated with a given $y^{(k-1)}$, this upper bound can become tighter than $B_{\rm max}$. Therefore, the final bit allocation in \eqref{eq:opt_sol_IA_QSMPA_1_1} reflects both the semantic importance and the delay tolerance constraints imposed by the system. This interaction between accuracy and latency effectively enables adaptive resource allocation that minimizes both quantization distortion and communication delay.

    After iterating from $k=1$ to $k=K$, a near-optimal solution $\left\{\bar{B}^{\star}[i], {P}^{\star}[i]\right\}_{i=1}^G$ and the associated latency $y^{\star}$ are obtained.
    Although the optimal number of quantization bits in \eqref{eq:opt_sol_IA_QSMPA_1_1} is derived as a real value, it must be an integer in practical implementations.
    To address this issue, we introduce a simple adjustment step that ensures integer quantization bits while fully utilizing the available bit and power budgets specified in \eqref{P_1_C_1} and \eqref{P_1_C_2}. 
    This step involves rounding and then clipping the solutions in \eqref{eq:opt_sol_IA_QSMPA_1_1}, expressed as $\tilde{B}[i] = {\rm Clipping}~({\rm Round}(\bar{B}^{\star}[i]), B_{\rm min}, B_{\rm max})$, followed by a refinement based on the residual bit gap $\Delta B = \sum_{i=1}^{G}D\tilde{B}[i] - B_{\rm target}$. 
    If $\Delta B < 0$ and $\tilde{B}[i] < B_{\rm max}$, one bit is incrementally added to the patch (i.e., $\tilde{B}[i] \gets \tilde{B}[i] + 1$) in descending order of mean attention scores. 
    Conversely, if $\Delta B > 0$ and $\tilde{B}[i] > B_{\rm min}$, one bit is removed (i.e., $\tilde{B}[i] \gets \tilde{B}[i] - 1$) in ascending order of mean attention scores. This adjustment continues until the total bit constraint in \eqref{P_1_C_1} is satisfied. 
    Finally, the refined bit allocation ${B^\star}[i] \in \left\{B_{\rm min}, \ldots, B_{\rm max}\right\}$  and the corresponding power allocation $P^\star[i]$, are applied to perform quantization and power control. 

    \subsection{Low-Complexity IA-QSMPA}\label{Sec:Low_complexity_EF}
    
    A major limitation of the IA-QSMPA method described in Sec.~\ref{Sec:IA_QSMPA_error_free}-B lies in its substantial computational complexity, quantified as $\mathcal{O}\left(KG \left(S_y  + S_{\nu}\right) + N_sF \log_2 (N_sF)\right)$, where $S_y$ and $S_{\nu}$ denote the maximum number of bisection steps required to determine $y$ and $\nu$, respectively. 
    Notably, computing $y^{(k)}$ requires a bisection process to solve a nonlinear system, resulting in a per-iteration complexity of $\mathcal{O}(S_y G)$. Similarly, the Lagrange multiplier $\nu^{\star (k)}$ is determined via bisection to meet the bit budget constraint in \eqref{eq:La_mul_1}, incurring an additional $\mathcal{O}(S_\nu G)$ complexity per iteration.
    Additionally, subcarrier mapping requires sorting all channel gains $\{\lambda_{f,r}\}_{\forall f,\forall r}$ contributing $\mathcal{O}(N_sF \log_2 (N_sF))$ operations.

    To alleviate this complexity, we propose a low-complexity variant of IA-QSMPA based on a separate optimization strategy, where quantization bit and power allocation are decoupled and optimized sequentially. In the bit allocation phase, we adopt the Top-$\beta$ strategy introduced in \cite{TF_IA_4}, where patches are ranked by their attention scores. The top $\beta\%$ of patches are assigned $B_{\rm max}$ bits, while the remaining patches are assigned $B_{\rm min}$ bits. This simple yet effective approach enables flexible control of the compression ratio with a single tunable parameter $k$. 
    Based on the Top-$\beta$ bit allocation $\{B^\star[i]\}_{i=1}^G$ and the derived equivalent channel gains in \eqref{eq:per_block_symbol_10_1}, the subsequent power allocation is formulated as the following convex optimization problem:
    \begin{align}
    ({\bf P}_3)~~&\underset{\{P[i]\}_{\forall i}}{\rm min}~ E_{\rm L}\big|_{B[i]=B^\star[i]}, \label{eq:P_3}\\
    {\rm s.t.}~~~
    &\eqref{P_1_C_2}, \eqref{P_1_C_4}. \notag
    \end{align}
    To solve the problem $({\bf P}_3)$, we leverage the analytical expressions already established for the optimal auxiliary variable and power control in Theorem~1. Specifically, by substituting the fixed bit allocation $\{B^\star[i]\}_{i=1}^G$ determined by the Top-$\beta$ strategy into the nonlinear equations \eqref{eq:dual_eq_1} and \eqref{eq:dual_eq_2}, the corresponding auxiliary variable $y$ and optimal power allocation $\{P^\star[i]\}_{i=1}^G$ can be obtained through a bisection-based dual optimization procedure.

    Although the proposed low-complexity method may yield a slight performance loss compared to the IA-QSMPA framework, it offers substantial complexity reduction. The Top-$\beta$ quantization and power allocation steps incur complexities of $\mathcal{O}(G)$ and $\mathcal{O}(S_y G)$, respectively, while subcarrier mapping adds $\mathcal{O}(N_sF\log_2 (N_sF))$. Consequently, the overall complexity is reduced to $\mathcal{O}(G(S_y+1) + N_sF\log_2 (N_sF))$, enabling efficient real-time deployment.

    \subsection{Inference Process of Overall System}\label{Sec:IA_QSMPA_overall}

    In the inference stage, the near-optimal solution $\left\{B^\star[i], P^\star[i] \right\}_{i=1}^{G}$ is employed. 
    Specifically, in the $t$-th OFDM symbol, the set of symbols $\{\bar{x}_{f,r}(t)\}_{\forall f \in \mathcal{F}_i, \forall r \in \mathcal{R}_i}$  share the same importance level, denoted by $I[i]$. These symbols are randomly mapped to the subchannels within the corresponding block $\left\{\lambda_{f,r}\right\}_{\forall f \in \mathcal{F}_i, \forall r \in \mathcal{R}_i}$.
    Under this subcarrier mapping, the achievable rate associated with the $i$-th block ${\bf x}^i$ can be expressed as
    \begin{align}\label{eq:IF_delay}
    R_{{\rm L}, i} &=\sum_{f \in {\mathcal F}_i}^{}\sum_{r \in {\mathcal R}_i}^{}\Delta f_0 \log_2\left(1+\frac{P_{f,r}(t)\lambda^2_{ f, r}}{\sigma^2}\right) \notag\\
    &\overset{(a)}{=} \sum_{f \in {\mathcal F}_i}^{}\sum_{r \in {\mathcal R}_i}^{}\Delta f_0 \log_2\left(1+\frac{P^\star[i]\lambda^2_{ f, r}}{\sigma^2}\right),
    \end{align}
    where $(a)$ follows from the fact that each subchannel within the $i$-th block is allocated the same transmission power (i.e., $P_{f,r}(t) = P^\star[i]$ for all $f \in \mathcal{F}_i$ and $r \in \mathcal{R}_i$).
    Given that the number of information bits transmitted in the $i$-th block is $DB[i]$, the overall worst-case communication latency during inference is determined as
    \begin{align}\label{eq:inf_latency}
        T_d = \underset{i}{\rm max}~\frac{DB[i]}{R_{{\rm L}, i}}.
    \end{align}

    \section{Modified IA-QSMPA Framework Under Finite Blocklength Transmission Scenario}\label{Sec:Modified_IA_QSMPA_error_robust}
    In this section, we extend the IA-QSMPA framework to practical MIMO-OFDM semantic communication systems, where semantic symbols are transmitted using finite blocklength coding, resulting in a nonzero probability of bit errors. To address this, we develop a transmission strategy that enhances semantic reliability by explicitly accounting for potential communication errors.
    
    \subsection{Finite Blocklength Transmission}\label{Sec:Modified_IA_QSMPA_1}

   In Sec.~\ref{Sec:IA_QSMPA_error_free}, we assumed reliable transmission of semantic symbols under the condition of sufficiently long blocklengths.
    However, in practical scenarios, the length of semantic symbols transmitted within each block is finite, and the probability of bit errors is nonzero. This necessitates a more realistic analysis that considers the fundamental limits of achievable rates under finite blocklength constraints.

    To address this issue, we analyze the impact of block error rate (BLER) on the achievable bit rate, providing a refined characterization of transmission performance under finite blocklength constraints.
    Given that $\lambda[i]$ is known from \eqref{eq:per_block_symbol_10_1}, the channel for each block can be equivalently modeled as an AWGN channel. In such channels, finite blocklengths impose a trade-off between the achievable bit rate and the block error probability, requiring the bit rate to be lowered to satisfy a target reliability constraint. Under these conditions, the upper bound of the achievable bit rate for the $i$-th block is given by \cite{Finite_block_cap}
    \begin{figure*}
    \begin{align}\label{eq:B_max_1}
        \bar{B}^{(k)}_{\rm max}[i] = 
        \underset{}{\rm min} \left(\frac{y^{(k-1)}\Delta f}{D}\left(\log_2\left(1+\gamma^{(k-1)}[i]\right) - \frac{{\sf Q}^{-1}(\tilde{\mu}_i)}{\ln 2\sqrt{L_c}}\sqrt{1-\frac{1}{\left(1+\gamma^{(k-1)}[i]\right)^2}} \right), B_{\rm max}\right).
    \end{align}
    \hrulefill	
    \end{figure*} 
    \begin{align}\label{eq:finite_length_cap}
        \!R_{\rm max}[i] &= \Delta f \left(\log_2(1+\gamma[i])-\frac{{\sf Q}^{-1}(\tilde{\mu}_i)}{\ln 2\sqrt{L_c}}\sqrt{U(\gamma[i])} \right), 
    \end{align}
    where $L_c=\frac{N_sFT}{G}$ denotes the maximum semantic symbol length across all blocks, satisfying $L[i] \leq L_c$ for all $i$, and $\tilde{\mu}_i$ represents the corresponding BLER. ${\sf Q}(x) = (2\pi)^{-1/2}\int_{x}^{\infty}e^{-t^2/2} dt$ denotes the Gaussian ${\sf Q}$-function, and  $U(\gamma[i]) = 1-1/(1+\gamma[i])^2$ represents the channel dispersion.
    Therefore, the worst-case communication latency can be defined as
    \begin{align}\label{eq:latency_FL}
         \bar{E}_{\rm L} = \underset{i}{\rm max}~\frac{DB[i]}{R_{\rm max}[i]}.
    \end{align}
    In this modification, we formulate a minimization problem that jointly considers weighted distortion and communication latency by incorporating the weighted quantization error $E_Q$ in \eqref{eq:grouping} and the communication latency in \eqref{eq:latency_FL}. The resulting bit and power allocation problem is given by
    \begin{subequations}
    \begin{align}
        ({\bf P}_4)~~&\underset{\{B[i], P[i]\}^{G}_{i=1}, y}{\rm min}~ y+{E}_{\rm Q}, \label{eq:P_4}\\
        {\rm s.t.}~~~ & \frac{DB[i]}{R_{\rm max}[i]} \leq y, \forall i \in \left\{1,\ldots, G\right\}, \label{P_4_C_1}\\
        &\eqref{P_1_C_1}\text{-}\eqref{P_1_C_4}.  \notag
    \end{align}
    \end{subequations}
    In the problem $({\bf P}_4)$, the use of the BCD algorithm does not guarantee convergence. Specifically, although the subproblem is convex with respect to $\{B[i]\}_{i=1}^G$ when $\{P[i]\}_{i=1}^G$ and $y$ are fixed, it becomes nonconvex with respect to $\{P[i]\}_{i=1}^G$ and $y$ for fixed $\{B[i]\}_{i=1}^G$, due to the nonlinear channel dispersion term in the latency constraint \eqref{P_4_C_1}.
    To address this issue, we approximate the channel dispersion term as a segment-wise linear function of $\gamma[i]$:
    \begin{align}\label{ch_disper}
        \sqrt{U(\gamma[i])} \approx \underset{}{\rm min}\left(\frac{\gamma[i]+1}{2}, 1\right).
    \end{align}
    Note that the linear term $\frac{\gamma[i]+1}{2}$ represents the tangent line to the function $\sqrt{U(\gamma[i])}$ at the point $\gamma[i]=0.4142$.
    \subsection{Modified Patch-Wise Quantization and Power Allocation}\label{Sec:Modified_IA_QSMPA_2}
        Utilizing the linearized approximation of the channel dispersion term, the communication latency constraint in \eqref{P_4_C_1} can be rewritten as
        \begin{align} \label{lat_1}
            &DB[i] \leq yR_{\rm max}[i] \notag \\
            &\overset{(a)}{\leq} y\Delta f \left\{\frac{\gamma[i]}{\ln 2} -\frac{2(1-\alpha_i)}{\ln 2}\sqrt{U(\gamma[i])}\right\}\notag \\
            &\overset{(b)}{\approx} y \underbrace{\Delta f\left\{\frac{\gamma[i]}{\ln 2} -\frac{2(1-\alpha_i)}{\ln 2}\underset{}{\rm min}\left(\frac{{\gamma}[i]+1}{2}, 1\right)\right\}}_{\triangleq \hat{R}_{\rm max}[i]},
        \end{align}
        where $\alpha_i = 1 - \frac{0.5{\sf Q}^{-1}(\tilde{\mu}_i)}{\sqrt{L_c}}$ and (a) follows from the inequality $\log_2(1+\gamma[i]) \leq \frac{\gamma[i]}{\ln 2}$, which is derived using the first-order Taylor approximation. (b) is based on the approximation given in \eqref{ch_disper}.
        By applying continuous relaxation to $B[i]$ under the constraint in \eqref{P_1_C_3} and incorporating the result of \eqref{lat_1}, the problem $({\bf P}_4)$ can be reformulated as
        \begin{subequations}
        \begin{align}
            ({\bf P}_5)~~&\underset{\{B[i], P[i]\}^{G}_{i=1}, y}{\rm min}~ y + {E}_{\rm Q}, \label{eq:P_5}\\
            {\rm s.t.}~~~ &DB[i] \leq  y\hat{R}_{\rm max}[i], \forall i \in \left\{1,\ldots, G\right\}, \label{P_5_C_1}\\
            &\eqref{P_1_C_1}, \eqref{P_1_C_2}, \eqref{P_2_C_2}. \notag
        \end{align}
        \end{subequations}
        To solve the nonconvex problem $({\bf P}_5)$, we apply the BCD algorithm following the same procedure as in Sec.~\ref{Sec:IA_QSMPA_error_free}-B. At each iteration, the optimal solution for each variable is sequentially derived by leveraging the KKT conditions. This procedure yields a near-optimal solution to $({\bf P}_5)$, as stated in the following theorem:
        \vspace{1mm}
        \begin{thm}
            Given that $\bar{B}^{(k)}_{\rm max}[i]$ is defined in \eqref{eq:B_max_1} and $\hat{B}^{(k)}[i]$ is obtained from \eqref{eq:opt_sol_IA_QSMPA_2_2}, the optimal solution of the problem $({\bf P}_5)$ in iteration $k$ of the BCD 
            algorithm is
        \begin{align}
             B^{(k)}[i] &= {\rm min}\left\{\bar{B}^{(k)}_{\rm max}[i], {\rm max}\left\{B_{\rm min}, \hat{B}^{(k)}[i] \right\}\!\right\}, \label{B_i_ER}
        \end{align}
        for all $i \in \{1,\ldots, G\}$, $k \in \{1,\ldots, K\}$.
        The optimal Lagrange multiplier $\nu^{\star (k)}$ is determined to satisfy the following equality:
        \begin{align}
            \sum_{i=1}^{G}B^{(k)}[i] = \frac{B_{\rm target}}{D}. \label{eq:B_k_i_sum}
        \end{align}
        In addition, the optimal slack variable $\tau^{\star (k)}$ is obtained by solving the nonlinear equation:
        \begin{align}
            \sum_{i=1}^{G}P^{(k)}[i] = \frac{P_{\rm tot}G}{N_sF}, \label{eq:P_k_i_sum}
        \end{align}
        where 
        {\small \begin{align}
        P^{(k)}[i]= \begin{cases}
            \frac{\sigma^2\ln 2}{\alpha_i \lambda^2[i]}\left\{\frac{D\Theta\Phi^{(k)}B^{(k)}[i]}{\Delta f \sqrt{\tau^{\star (k)}}}+\frac{1-\alpha_i}{\ln 2}\right\},~{\rm if} ~\gamma^{(k-1)}[i] < 1, \label{eq:U_i_1}  \\
            \frac{\sigma^2\ln 2}{\lambda^2[i]}\left\{\frac{D\Theta\Xi^{(k)}B^{(k)}[i]}{\Delta f \sqrt{\tau^{\star (k)}}}+\frac{2(1-\alpha_i)}{\ln 2}\right\}, ~\text{otherwise}, \end{cases} 
        \end{align}}for all $i \in \{1,\ldots, G\}$, $k \in \{1,\ldots, K\}$. Here,   
        \begin{align*}
            \Theta &= \sqrt{\frac{\Delta f_0}{\sigma^2 D \ln 2}}, \\
            \Phi^{(k)} &= \frac{1}{\sqrt{\sum_{j=1}^{G} \frac{B^{(k)}[j]}{\alpha_j \lambda^2[j]}}}, \\
            \Xi^{(k)} &= \frac{1}{\sqrt{\sum_{j=1}^{G} \frac{B^{(k)}[j]}{\lambda^2[j]}}}.
        \end{align*}
        Once $\{B^{(k)}[i]\}_{i=1}^{G}$ and $\tau^{\star(k)}$ are determined, the associated auxiliary variable $y^{(k)}$ can be computed as
        \begin{align}
             y^{(k)} &= \begin{cases}
            \frac{\sqrt{\tau^{\star (k)}}}{\Theta\Phi^{(k)}},~{\rm if} ~{\gamma}^{(k-1)}[i] < 1, \label{y_k} \\
            \frac{\sqrt{\tau^{\star (k)}}}{\Theta\Xi^{(k)}}, ~\text{otherwise}. \end{cases}
        \end{align}
        \end{thm}
    \begin{IEEEproof}
         See Appendix B.
    \end{IEEEproof}
    \vspace{1mm}
    
    In the ideal transmission scenario discussed in Sec.~{\ref{Sec:IA_QSMPA_error_free}-D}, the worst-case communication latency is evaluated based on ideal achievable rates assuming infinite blocklength. In contrast, under the finite blocklength regime, the rate for each symbol in the $i$-th block must be adjusted to account for the reliability constraint imposed by a non-negligible $\tilde{\mu}_i$.
    Accordingly, the effective achievable rate for transmitting the semantic symbols in the $i$-th block ${\bf x}^i$ is given by
    \begin{align}\label{eq:finite_rate_main}
    R_{{\rm L}, i} = \sum_{f \in \mathcal{F}_i} \sum_{r \in \mathcal{R}_i} \max\left(C_{f,r}(t) - \Gamma_{f,r}(t), 0\right),
    \end{align}
    where $C_{f,r}(t) = \Delta f_0\log_2\left(1+\gamma_{f,r}(t)\right)$ and $\Gamma_{f,r}(t) = \Delta f_0\frac{2(1-\alpha_i)}{\ln 2}\sqrt{1-\frac{1}{\left(1+\gamma_{f,r}(t)\right)^2}}$, with $\gamma_{f,r}(t) = \frac{P_{f,r}(t)\lambda_{f,r}^2}{\sigma^2}$. 
    As in the ideal transmission case, all subchannels within the $i$-th block are allocated the same transmission power, and hence $\gamma_{f,r}(t)$ can be expressed as a function of $P^\star[i]$.
    The $\max(\cdot,0)$ operator ensures that the effective achievable rate remains non-negative even under severe channel conditions. Using this definition, the worst-case communication latency $T_d$ during inference is expressed as in \eqref{eq:inf_latency}.
    

    \subsection{Modified Low-Complexity IA-QSMPA}\label{Sec:Low_complexity_ER}

    We also extend the low-complexity strategy presented in Sec.~\ref{Sec:IA_QSMPA_error_free}-C by incorporating the effect of non-zero BLER under finite blocklength constraints.
    Based on a given Top-$\beta$ bit allocation $\{B^\star[i]\}_{i=1}^G$, the corresponding power allocation problem is formulated as 
    \begin{align}
        ({\bf P}_6)~~&\underset{\{P[i]\}_{\forall i}}{\rm min}~ \bar{E}_{\rm L}\big|_{B[i]=B^\star[i]}, \label{eq:P_6}\\
        {\rm s.t.}~~~
        &\eqref{P_1_C_2}, \eqref{P_1_C_4}. \notag
    \end{align}
    Although the original formulation of $\bar{E}_{\rm L}$  is nonconvex due to the presence of channel dispersion terms, it becomes convex once these terms are appropriately linearized. Exploiting this convexified structure, the KKT conditions yield the following piecewise expression for the optimal power allocation:
    \begin{align}
        \!\! P^\star[i] &= \begin{cases}
        \frac{\sigma^2}{\lambda^2[i]}\bar{\gamma}[i],~{\rm if} ~\bar{\gamma}[i] < 1, \label{eq:LC_P_opt}  \\
        \frac{\sigma^2\ln 2}{\lambda^2[i]}\left\{\frac{DB^\star[i]\Theta\Xi}{\Delta f \sqrt{\tau^{\star }}}+\frac{2(1-\alpha_i)}{\ln 2}\right\}, ~\text{otherwise}, 
        \end{cases} 
    \end{align}
    where $\bar{\gamma}[i] = \frac{\ln 2}{\alpha_i}\left\{\frac{D B^\star[i]\Theta\Phi}{\Delta f\sqrt{\tau^{\star}}}+\frac{(1-\alpha_i)}{\ln 2}\right\}$. The terms $\Phi$ and $\Xi$ are obtained by substituting $B^{(k)}[i]$ with $B^\star[i]$ in the corresponding expressions for $\Phi^{(k)}$ and $\Xi^{(k)}$, respectively. The optimal Lagrange multiplier $\tau^\star$ is determined to satisfy the total power constraint (i.e., $\sum_{i}P^\star[i] = \frac{P_{\rm tot}G}{N_sF}$). The resulting expression in \eqref{eq:LC_P_opt} closely follows the structure of \eqref{eq:U_i_1}, with the key distinction that it provides a closed-form solution obtained in a single step, rather than through iterative updates within the BCD algorithm.

   The computational complexity of the modified IA-QSMPA framework is derived in the same manner as the IA-QSMPA method in Sec.~\ref{Sec:IA_QSMPA_error_free}-C, resulting in a complexity of $\mathcal{O}\left(KG \left(S_\tau  + S_{\nu}\right) + N_sF\log_2 (N_sF)\right)$, where $S_\tau$ denotes the maximum number of bisection steps required to determine $\tau$.
   In contrast, the modified low-complexity method reduces this to $\mathcal{O}\left(G (S_\tau + 1) + N_sF \log_2 (N_sF)\right)$, providing a more efficient solution while maintaining robustness under finite blocklength transmission scenarios.

   \section{Simulation Results}\label{Sec:Simul}
   In this section, we evaluate the superiority of the proposed IA-QSMPA methods through simulations. In these simulations, we consider the following task:
   \begin{itemize}
       \item {\bf Multi-view image classification:}
        We evaluate multi-view image classification using the MVP-N \cite{MVP_N} dataset. In this setting, four devices transmit images of the same object from different viewpoints. The server receives one image per device, classifies each image individually, and determines the final prediction through majority voting.
        
   \end{itemize}

    All input images are resized to $(3, 224, 224)$ and normalized to have zero mean and unit variance. The ViT encoder deployed on the device is DeiT-Tiny, while DeiT-Small is used on the server, featuring approximately 4.4 times more parameters~\cite{D_model}. 
    Both models are pretrained on ImageNet-1k and share the same architectural parameters ($P=16$, $L=12$, $N=196$).
    The device model is configured with a hidden dimension of $D=192$ and $H=3$ attention heads, while the server model adopts $D=384$ and $H=6$. For classification, the class token is passed through a fully connected layer. Fine-tuning is performed offline using cross-entropy loss, the Adam optimizer with a learning rate of 0.0001, a batch size of 32, and 10 training epochs. This process is conducted independently of communication protocols or channel modeling.
   All configurations are based on $N_{\rm tx} = N_{\rm rx} = N_s \in \{4,6,8\}$, $F = 784$, $T = 50$, $\Delta f_0 = 15~{\rm kHz}$, $\sigma_H^2 = 1$, $G = 196$, and $P_{\rm tot} = 3136$.
   The compression ratio is defined as the ratio of the compressed bit overhead to the original bit overhead (i.e., $\rho = \frac{B_{\rm target}}{8HWC}$). 
   To avoid dependency on specific channel coding and modulation schemes, we assume that in the ideal transmission scenario, the transmitted signal is reliably delivered. In the finite blocklength scenario, transmission is carried out under a predefined reliability constraint, such as a target BLER $\tilde{\mu}_i$. To account for bit-level reliability, we estimate the corresponding bit error rate, which can be approximated as
   \begin{align}
        {\rm BER}_i = \gamma\left\{1-(1-{\tilde{\mu}_i})^{\frac{1}{L_c}}\right\},
    \end{align}
   where $\gamma$ is a correction factor set as $\gamma=10$.

   For performance comparison, we consider the following methods:

   \begin{itemize}
        \item {\bf IA-QSMPA}: This is the proposed method for the ideal transmission scenario, which employs IASM and joint bit-power allocation procedure described in Sec.~\ref{Sec:IA_QSMPA_error_free}-B to solve the optimization problem $({\bf P}_2)$.
        \item {\bf IA-QSMPA (LC)}: This is a low-complexity alternative of {\bf IA-QSMPA}, described in Sec.~\ref{Sec:IA_QSMPA_error_free}-C, which decouples bit and power allocation and solves the simplified problem $({\bf P}_3)$.
        \item {\bf Modified IA-QSMPA}: This is the proposed method for the finite blocklength transmission scenario, which applies the optimization strategy detailed in Sec.~\ref{Sec:Modified_IA_QSMPA_error_robust}-B to solve the problem $({\bf P}_5)$.
        \item {\bf Modified IA-QSMPA (LC)}: This is a low-complexity alternative of {\bf Modified IA-QSMPA}, described in Sec.~\ref{Sec:Modified_IA_QSMPA_error_robust}-C, which separately optimizes bit and power allocation based on the problem $({\bf P}_6)$. 

       \begin{figure}[t]
        \centering   
        {\epsfig{file=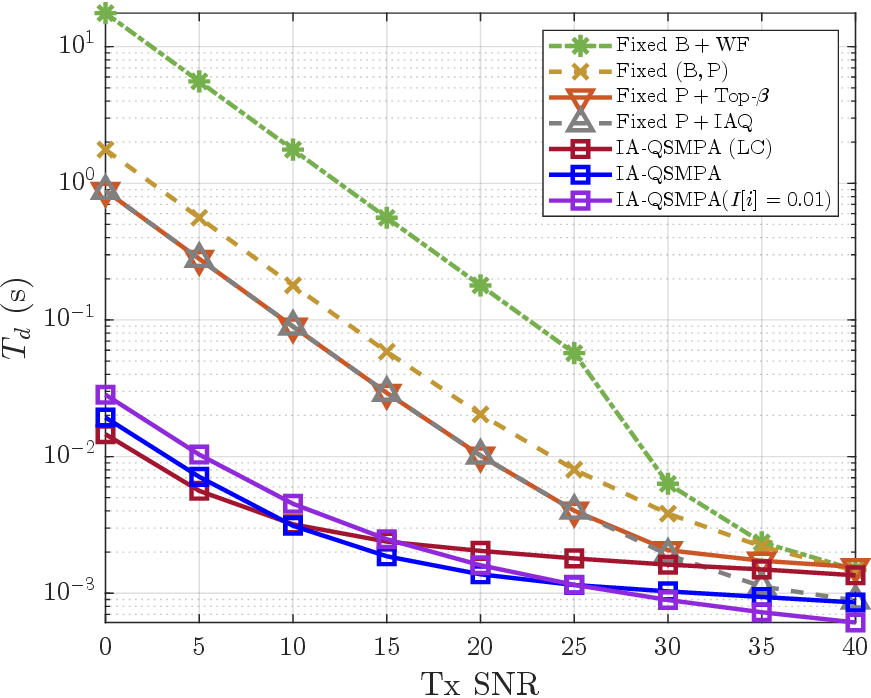, width=7.2cm}}
        \caption{Comparison of the worst-case communication latency across various transmission methods for a multi-view image classification task on the MVP-N dataset when $N_s=4$ and $\rho=0.25$.}
        \label{fig:EF_T_d}
    \end{figure}
  
    \item {\bf Fixed ({B, P})}: This is a fixed-level quantization and power allocation method, where both the quantization bit $B[i]$ and the transmission power $P[i]$ are uniformly assigned across all blocks. In this setting, the compression ratio simplifies to 
    $\rho = \frac{GDB[i]}{8HWC}$, as $B[i]$ remains constant for all $i$.    

    \item {\bf Fixed B + WF}: This is a fixed-level quantization method combined with WF power allocation. In this setting, the quantization bit $B[i]$ is uniformly assigned across all blocks, while the transmission power $P[i]$ is allocated based on the classical WF solution optimized for SVD beamforming in conventional MIMO-OFDM systems.


    \item {\bf Fixed P + {\bf IAQ}}:
    This is the IAQ method proposed in \cite{TF_IA_5}. In this setting, the quantization level for each patch is determined based on its semantic importance, while the transmission power is uniformly assigned across all patches.

    \item {\bf Fixed P + {\bf Top-$\bm \beta$}}:
    This is the attention-aware patch selection method proposed in \cite{TF_IA_4}, as detailed in Sec.~\ref{Sec:Low_complexity_EF}.
    The compression ratio simplifies to 
    \begin{align}
        \rho = \frac{GD \left(kB_{\rm max}+(100-k)B_{\rm min}\right)}{100\times 8HWC},
    \end{align}
    and the transmission power is uniformly allocated across all patches.



   \end{itemize}

   For all the proposed methods, we set  $K=5$, $B_{\rm min}=1$, and $B_{\rm max}=8$. 
   The importance weight $I[i]$ is defined as
    \begin{align}\label{eq:weight_fn}
        I[i] = \frac{1-d_c}{(a_{\rm max}-a_{\rm min})^\delta}(a_i-a_{\rm min})^\delta+d_c,
    \end{align}
    where $a_{\rm min} = \underset{i}{\rm min}(a_i)$, $a_{\rm max} = \underset{i}{\rm max}(a_i)$. The exponent $\delta >0$ controls the sharpness of the resulting weight distribution and set as 1. and A small constant $d_c=10^{-7}$ is added to prevent the weight from becoming zero.

   Notably, when all weights are uniformly set to a constant value strictly less than one (e.g., $I[i] = 0.01, \forall i$), the bit allocation becomes uniform across blocks. In this case, only the power allocation needs to be optimized.

        \begin{figure}[t]
        \begin{minipage}{1\columnwidth}
            \centering
            \subfigure[${\rm Tx~SNR}=20~{\rm dB}$, $N_s=N_{\rm tx}=N_{\rm rx}=4$]
            {\epsfig{file=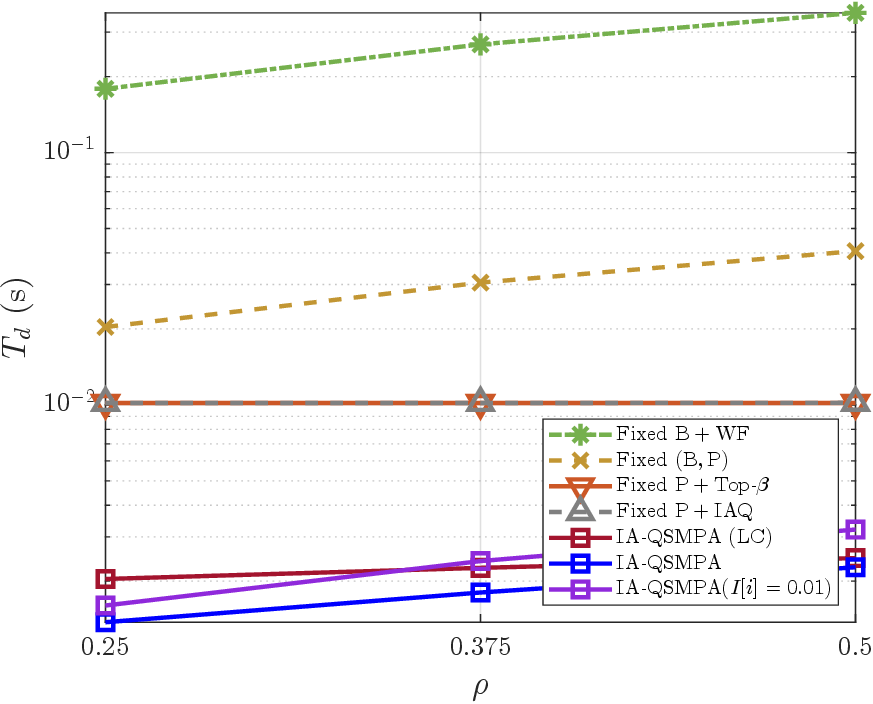, width=7.2cm}}
            \hspace{3mm}
            \subfigure[${\rm Tx~SNR}=20~{\rm dB}$, $N_s=N_{\rm tx}=N_{\rm rx}=4$]
		{\epsfig{file=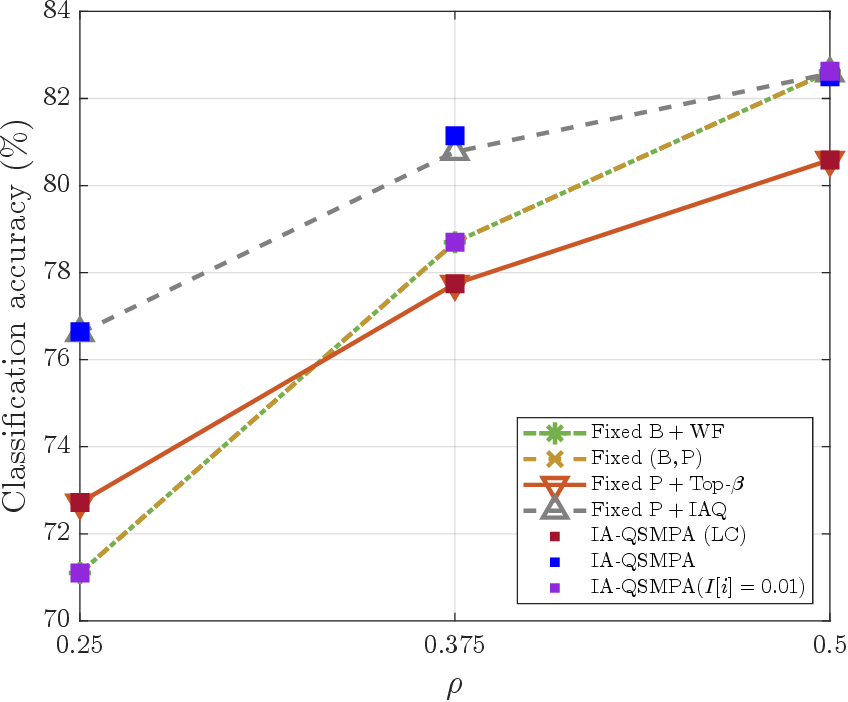, width=7.2cm}} 
            \captionof{figure}{
            Comparison of the worst-case communication latency and classification accuracy across various transmission methods under different $\rho$ in a multi-view image classification task on the MVP-N dataset.}
           \label{fig:EF_T_d_Acc}
        \end{minipage}
    \end{figure}
    
    \subsection{Performance Evaluation under Ideal Transmission Scenario}

    Fig.~\ref{fig:EF_T_d} compares the worst-case communication latency ($T_d$) across various transmission methods for the multi-view image classification task on the MVP-N dataset.
    Fig.~\ref{fig:EF_T_d}  shows that {\bf IA-QSMPA} consistently achieves the lowest communication latency among all methods. This improvement is attributed to its joint optimization of subcarrier mapping and bit-power allocation, which enables the system to concentrate resources on semantically important regions.
    {\bf IA-QSMPA (LC)} also demonstrates strong performance, outperforming all baseline methods when ${\rm Tx~SNR} \leq 30~{\rm dB}$. Although some degradation is observed at higher SNRs due to its decoupled optimization structure, it remains a competitive alternative with favorable latency performance.
    In addition, {\bf IA-QSMPA ($I[i]=0.01$)}, which assigns uniform importance weights to all patches, shows degraded performance in the low SNR regime due to its limited ability to reflect semantic importance. However, as the objective shifts toward latency reduction, it achieves a sharp decrease in $T_d$ at high SNRs, demonstrating effectiveness in delay-sensitive scenarios despite the lack of importance adaptivity.
    Meanwhile, {\bf Fixed B + {\bf WF}} exhibits inferior performance, particularly when compared to {\bf Fixed ({B, P})}. This suggests that uniform power allocation, rather than the conventional WF strategy, is more effective in minimizing the worst-case latency.
    Moreover, under uniform power allocation, {\bf Fixed P + {\bf Top-$\bm \beta$}} and {\bf Fixed P + {\bf IAQ}} achieve lower latency than {\bf Fixed ({B, P})} by assigning only 1 bit to the least important patches, which dominate the overall delay. In contrast, {\bf Fixed ({B, P})} assigns 2 bits uniformly to all patches, resulting in longer transmission time.
    
    Fig.~\ref{fig:EF_T_d_Acc} compares the worst-case communication latency ($T_d$) and classification accuracy across various transmission methods under different compression ratios $\rho$ for the multi-view image classification task on the MVP-N dataset.
    Fig.~\ref{fig:EF_T_d_Acc}(a) shows that {\bf IA-QSMPA} consistently achieves the lowest transmission delay across all values of $\rho$, and {\bf IA-QSMPA (LC)} also outperforms the fixed power and fixed bit schemes.
    Fig.~\ref{fig:EF_T_d_Acc}(b) further shows that {\bf IA-QSMPA} attains the highest classification accuracy among all methods, demonstrating the effectiveness of its IAQ strategy.
    Moreover, {\bf IA-QSMPA ($I[i]=0.01$)}, which assigns the same number of quantization bits to all blocks by setting uniform importance weights, achieves performance comparable to the fixed bit schemes.
    Furthermore, compared to binary quantization strategies such as {\bf Top-$\bm \beta$}, the finer-grained quantization used in {\bf IAQ} and {\bf IA-QSMPA} results in substantial improvements in classification accuracy.


    \begin{figure}[t]
        \centering   
        {\epsfig{file=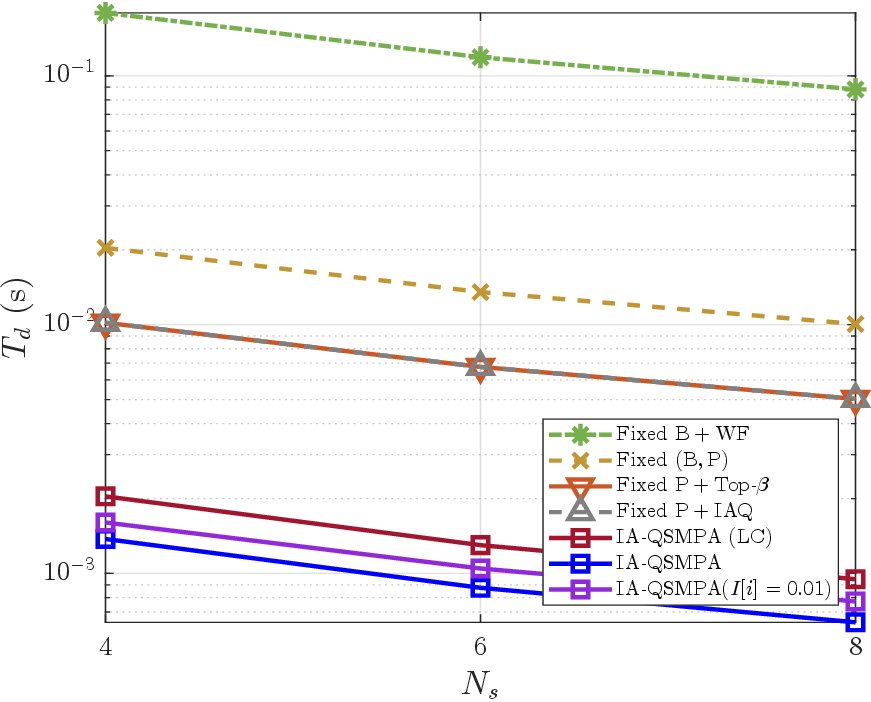, width=7.2cm}}
        \caption{Comparison of the worst-case communication latency across various transmission methods under different $N_s$ for a multi-view image classification task on the MVP-N dataset when $\rho=0.25$ and ${\rm Tx~SNR}=20~{\rm dB}$.}
        \label{fig:EF_T_d_N_s}
    \end{figure}

        \begin{figure*}
        \centering
        \begin{minipage}{2\columnwidth}
            \centering
            \subfigure[$\tilde{\mu}_i = 0.01$]
            {\epsfig{file=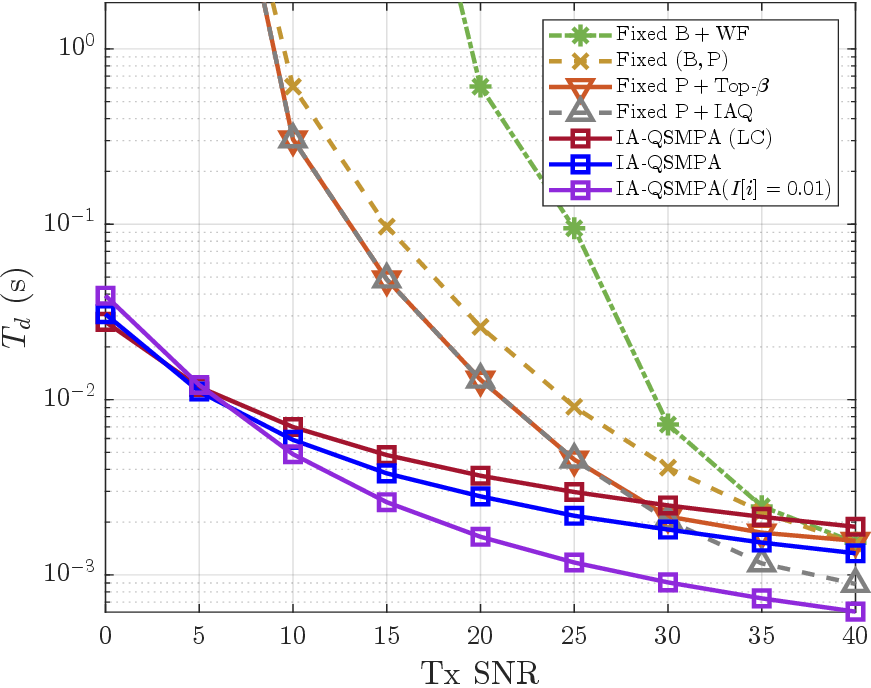, width=6cm}}
            \subfigure[$\tilde{\mu}_i = 0.1$]
		  {\epsfig{file=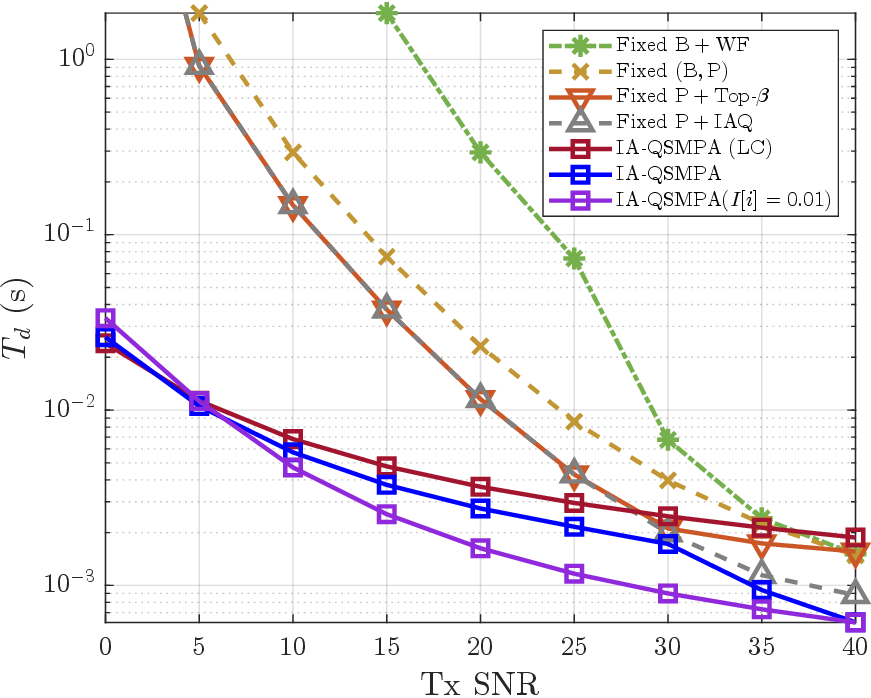, width=6cm}}
            \subfigure[$\tilde{\mu}_i = 0.1, {\rm Tx~SNR} = 20~{\rm dB}$]
		  {\epsfig{file=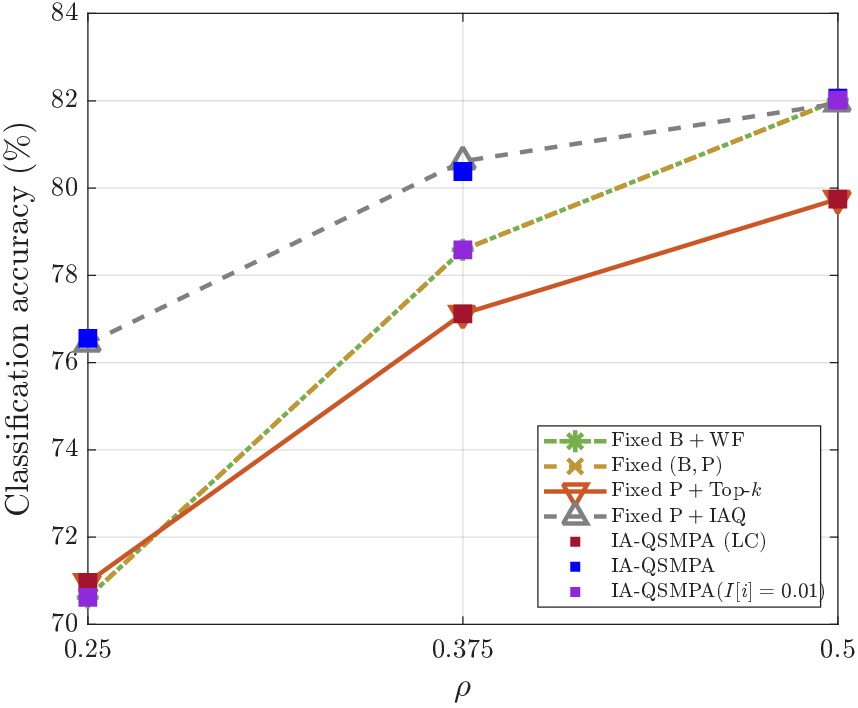, width=6cm}}
            \captionof{figure}{Comparison of the worst-case communication latency and classification accuracy across various transmission methods under finite blocklength transmission for a multi-view image classification task on the MVP-N dataset at $N_s = 4$, and $\rho = 0.25$.}
            \label{fig:ER_T_d}
        \end{minipage}
    \end{figure*}
    Fig.~\ref{fig:EF_T_d_N_s} compares the worst-case communication latency ($T_d$) across various transmission methods under different $N_s$  for the multi-view image classification task on the MVP-N dataset.
    As $N_s$ increases, more subchannels are allocated to each block, leading to higher achievable rates per block according to \eqref{eq:IF_delay}.
    As a result, all methods exhibit a monotonic decrease in $T_d$.
    Notably, {\bf IA-QSMPA} consistently achieves the lowest latency under all spatial configurations, demonstrating its robustness and effectiveness in leveraging additional spatial resources. Although the results are not explicitly plotted, in the above simulations, we confirm that {\bf IA-QSMPA} maintains the highest classification accuracy across all values of $N_s$, confirming its superior performance in terms of both communication efficiency and task performance.

    \subsection{Performance Evaluation under Finite Blocklength Transmission Scenario}
    


   Fig.~\ref{fig:ER_T_d} compares the worst-case communication latency ($T_d$) and classification accuracy across various transmission methods under the finite blocklength transmission scenario for the multi-view image classification task on the MVP-N dataset. 
    Fig.~\ref{fig:ER_T_d}(a) shows that latency increases significantly in the low SNR regime compared to Fig.~\ref{fig:EF_T_d}, primarily due to the reduced achievable rate caused by finite blocklength effects discussed in Sec.~\ref{Sec:Modified_IA_QSMPA_error_robust}.
    At high SNR, {\bf IA-QSMPA} shows a slight latency increase relative to {\bf Fixed P + IAQ}, primarily due to the large impact of approximation errors in the piecewise linear model in \eqref{ch_disper}.
    Notably, {\bf IA-QSMPA ($I[i] = 0.01$)} achieves lower latency than {\bf IA-QSMPA} when ${\rm Tx~SNR} > 5~{\rm dB}$, showing a similar tendency to that observed in Fig.~\ref{fig:EF_T_d}. However, as shown in Fig.~\ref{fig:ER_T_d}(c), this latency benefit comes at the expense of reduced classification accuracy, revealing a trade-off between communication efficiency and task performance.
    Additionally, Fig.~\ref{fig:ER_T_d}(b) shows that the proposed methods maintain stable latency performance in the low SNR regime, even as $\tilde{\mu}_i$ increases. In contrast, the baseline methods exhibit noticeable degradation, highlighting the superior robustness of the proposed methods against decoding errors.

     \begin{table}[t] \renewcommand{\arraystretch}{1.2} 
        \caption{Classification accuracy of different subcarrier mapping strategies under fixed bit and fixed power configurations in an uncoded MIMO-OFDM system using 64-QAM with $\rho=0.5$.} \label{table:uncoded_ia_sm} 
        \setlength{\tabcolsep}{3pt} 
        \footnotesize 
        \centering 
        \begin{tabular}{|c|c|c|c|c|} 
        \hline 
        Tx SNR & 0~dB & 10~dB & 20~dB & 30~dB  \\ \hline 
        \bf{Fixed ({B, P}) + Inverse SM} & $14.55$ & $48.02$ & $67.89$  & $77.55$ \\ \hline 
        \bf{Fixed ({B, P}) + Random SM} & $29.38$ & $63.03$ & $74.98$ & $80.94$  \\ \hline
        \bf{Fixed ({B, P}) + IASM} & $55.59$ & $76.45$ & $81.95$ & $82.51$  \\ \hline 
        \end{tabular} 
    \end{table}

    Table~\ref{table:uncoded_ia_sm} presents the classification accuracy of three subcarrier mapping strategies under fixed bit and fixed power configurations in an uncoded MIMO-OFDM system using 64-ary quadrature amplitude modulation (64-QAM). Inverse subcarrier mapping ({\bf Inverse SM}) assigns better channels to less important patches, while random subcarrier mapping ({\bf Random SM}) allocates subcarriers without considering semantic importance.
    The results demonstrate that {\bf IASM} consistently outperforms  other subcarrier mapping strategies across all SNR levels when comparing schemes with identical power allocation. These findings highlight the advantage of {\bf IASM}, which prioritizes the reliable delivery of semantically important information by leveraging stronger subcarriers. 

   \section{Conclusion}
   In this paper, we proposed IA-QSMPA, an importance-aware semantic communication framework designed for MIMO-OFDM systems. The framework jointly optimizes quantization levels, subcarrier mapping, and power allocation by leveraging semantic importance scores derived from a pretrained ViT model. This importance-guided design enables improved task performance and transmission efficiency. We first introduced the IA-QSMPA framework under ideal transmission conditions and then extended it to more practical settings involving finite blocklength transmissions, where the achievable rate is limited by a nonzero decoding error probability. Through extensive simulations on a multi-view image classification task, we demonstrated that IA-QSMPA consistently outperforms existing transmission schemes in terms of both semantic task accuracy and communication efficiency. These results highlight the potential of semantic importance-driven design in advancing the robustness and adaptability of semantic communication systems in realistic wireless systems.
   

    A promising direction for future research is to extend IA-QSMPA to accommodate dynamic channel conditions with imperfect channel state information. Another potential direction is to apply the proposed framework to multi-user semantic communication systems.

    \appendices
    \section{Proof of Theorem 1}\label{apdx:thm_1}

    This appendix presents the derivation of the BCD algorithm for solving the problem (${\bf P}_2$), as introduced in Sec.~\ref{Sec:IA_QSMPA_error_free}-B. The optimization procedure alternately updates the variables $\{B^{(k)}[i]\}_{i=1}^G$, $\{P^{(k)}[i]\}_{i=1}^G$, and $y^{(k)}$ at each iteration $k$ to efficiently solve the the problem $({\bf P}_2)$.
    To simplify the optimization structure, the two constraints in \eqref{P_2_C_1} and \eqref{P_2_C_2} are first reformulated into a single equivalent condition by introducing an upper bound on $B[i]$, as shown in \eqref{eq:opt_sol_IA_QSMPA_2_1}. This leads to a compact range constraint, i.e., 
    \begin{align}\label{eq:append_A_1}
    B_{\rm min} \leq B^{(k)}[i] \leq \bar{B}^{(k)}_{\rm max}[i], 
    \end{align} 
    for all $i \in \{1,\ldots,G\}$. Given $\{P^{(k-1)}[i]\}_{i=1}^G$ and $y^{(k-1)}$, the subproblem with respect to $\{B^{(k)}[i]\}_{i=1}^G$ becomes convex, as ensured by \eqref{P_1_C_1} and \eqref{eq:append_A_1}. In this case, following the methodology proposed in \cite{TF_IA_5}, the optimal values of $\{B^{(k)}[i]\}_{i=1}^G$ at iteration $k$ are obtained in closed form, as given in \eqref{eq:opt_sol_IA_QSMPA_1_1} \eqref{eq:opt_sol_IA_QSMPA_2_1}, 
    \eqref{eq:opt_sol_IA_QSMPA_2_2}, and
    \eqref{eq:La_mul_1}.

    Once $\{B^{(k)}[i]\}_{i=1}^G$ is fixed, the problem $({\bf P}_2)$ reduces to the following convex subproblem with respect to the auxiliary variable $y^{(k)}$ and the power allocation $\{P^{(k)}[i]\}_{i=1}^{G}$:
    \begin{subequations}
    \begin{align}
        &({\bf P}_2\text{-}1)~~\underset{\{P^{(k)}[i]\}_{i=1}^{G}, y^{(k)}}{\rm min}~ y^{(k)} \label{eq:proof_IA_QSMPA_1_1}\\
        {\rm s.t.}~~
        &DB^{(k)}[i] \leq y^{(k)}\Delta f\log_2\left(1+\gamma^{(k)}[i]\right),~\forall i, \label{eq:proof_IA_QSMPA_1_2}\\
        &\sum_{i=1}^{G}\frac{N_sF}{G}P^{(k)}[i]=P_{\rm tot}. \label{eq:proof_IA_QSMPA_1_3} 
    \end{align}
    \end{subequations}
    To solve this subproblem, the KKT conditions are applied. Let $\tau^{ (k)}$ denote the Lagrange multiplier associated with the total power constraint, and $\rho^{(k)}_i \geq 0$ denote the Lagrange multipliers for the latency constraints. The corresponding Lagrangian is given by
    \begin{align}\label{eq:proof_IA_QSMPA_2}
    &\mathcal{L}(\tau^{ (k)}, \{\rho^{(k)}_i\}_{i=1}^{G}) \nonumber \\
    &= y^{(k)} + \tau^{ (k)}\left(\sum_{i=1}^{G}\frac{N_sF}{G}P^{(k)}[i] - P_{\rm tot}\right) \notag\\
    &~~~+ \sum_{i=1}^{G}\rho^{(k)}_i\left(\frac{DB^{(k)}[i]}{\Delta f \log_2\left(1 + \gamma^{(k)}[i]\right)} - y^{(k)}\right).
    \end{align}
    The stationarity conditions are obtained by differentiating $\mathcal{L}(\cdot)$ with respect to $P^{(k)}[i]$ and $y^{(k)}$, leading to
    \begin{align}
    \frac{\partial \mathcal{L}}{\partial P^{(k)}[i]} =&~ \tau^{(k)} \frac{N_s F}{G} - \rho^{(k)}_i \frac{DB^{(k)}[i] \lambda^2[i]}{\Delta f  \sigma^2 \ln 2} \nonumber \\
    &\times \frac{\left(1+\gamma^{(k)}[i]\right)^{-1}}{\left\{\log_2\left(1+\gamma^{(k)}[i]\right)\right\}^2} = 0, \forall i, \label{eq:proof_IA_QSMPA_3_1}\\
    \frac{\partial \mathcal{L}}{\partial y^{(k)}} =&~ 1 - \sum_{i=1}^{G}\rho^{(k)}_i = 0. \label{eq:proof_IA_QSMPA_3_2}
    \end{align}
    The complementary slackness condition imposes
    \begin{align}\label{eq:proof_IA_QSMPA_4}
        \frac{DB^{(k)}[i]}{\Delta f \log_2 (1+\gamma^{(k)}[i])} \leq y^{(k)}, \forall i.
    \end{align}
    This inequality must hold with equality for all $i$. If all $\rho^{(k)}_i$ were zero, then the normalization condition in \eqref{eq:proof_IA_QSMPA_3_2} would be violated, since it requires $\sum_i \rho^{(k)}_i = 1$. Therefore, at least one $\rho^{(k)}_i$ must be strictly positive. However, if some $\rho^{(k)}_i$ are positive while others are zero, then from the stationarity condition in \eqref{eq:proof_IA_QSMPA_3_1}, the resulting values of $\tau^{(k)}$ become inconsistent across $i$, leading to a contradiction. Consequently, it must hold that $\rho^{(k)}_i > 0$ for all $i$, which implies that the corresponding inequality constraints are active, which implies that the equality in \eqref{eq:proof_IA_QSMPA_4} must be satisfied. Under this condition, the expression for $\rho^{(k)}_i$ is given by
    \begin{align}\label{eq:proof_IA_QSMPA_6}
    \rho^{(k)}_i = \tau^{(k)} \ln 2 \cdot \frac{ \sigma^2 D B^{(k)}[i]}{\Delta f_0 \lambda^2[i] \{y^{(k)}\}^2} \cdot 2^{\frac{D B^{(k)}[i]}{y^{(k)}\Delta f}}.
    \end{align}
    {Substituting this relation into the normalization condition in \eqref{eq:proof_IA_QSMPA_3_2} yields \eqref{eq:dual_eq_1}.
    Applying the equality condition in \eqref{eq:proof_IA_QSMPA_4} and the total power constraint in \eqref{P_1_C_2} leads to \eqref{eq:dual_eq_2}.
    The optimal value of $y^{(k)}$ is then obtained via a bisection search on \eqref{eq:dual_eq_2}, and the corresponding transmit powers $\{P^{(k)}[i]\}_{i=1}^G$ are computed using \eqref{eq:opt_sol_IA_QSMPA_1_3}.}
    This completes the proof of Theorem~1.

    \section{Proof of Theorem 2}\label{apdx:thm_2}

    This appendix presents the derivation of the BCD algorithm for solving the problem $({\bf P}_5)$, as introduced in Sec.~\ref{Sec:Modified_IA_QSMPA_error_robust}-B. Given $\{P^{(k-1)}[i]\}_{i=1}^{G}$ and $y^{(k-1)}$ at iteration $k$ of the BCD algorithm, the subproblem with respect to $\{B^{(k)}[i]\}_{i=1}^{G}$ becomes convex. To improve the accuracy of the bit allocation step, the loose upper bound $\hat{R}_{\rm min}[i]$ in \eqref{P_5_C_1} is replaced with the tighter bound $R_{\rm min}[i]$ derived from \eqref{eq:finite_length_cap}. This substitution preserves the convexity of the subproblem and yields a refined expression for the bit allocation upper bound $\bar{B}^{(k)}_{\rm max}[i]$, as given in \eqref{eq:B_max_1}. The optimal bit allocation $\{B^{(k)}[i]\}_{i=1}^{G}$ is then computed in closed form using the procedure detailed in Appendix~A and \cite{TF_IA_5}, as expressed in \eqref{eq:opt_sol_IA_QSMPA_2_2}, \eqref{eq:B_max_1}, \eqref{B_i_ER}, and \eqref{eq:B_k_i_sum}.

    Once $\{B^{(k)}[i]\}_{i=1}^G$ is fixed, the problem $({\bf P}_5)$ reduces to the following convex subproblem with respect to the auxiliary variable $y^{(k)}$ and the power allocation $\{P^{(k)}[i]\}_{i=1}^{G}$:
    \begin{subequations}
    \begin{align}\label{eq:proof_Modified_IA_QSMPA_1}
        &({\bf P}_5\text{-}1)~~\underset{\{P^{(k)}[i]\}_{i=1}^{G}, y^{(k)}}{\rm min}~ y^{(k)} \\
        {\rm s.t.}~~
        &\frac{DB^{(k)}[i]\ln 2}{\Delta f\left(\gamma^{(k)}[i]-2(1-\alpha_i)\underset{}{\rm min}\left(\frac{{\gamma}^{(k)}[i]+1}{2}, 1\right) \right)} \leq y^{(k)},~\forall i, \\
        &\eqref{eq:P_k_i_sum}. \notag
    \end{align}
    \end{subequations}
    To solve this subproblem, we consider the following two cases:
    \begin{itemize}
        \item[(i)] {\bf Case 1} ($0 \leq \gamma^{(k-1)}[i]<1$): In this case, the Lagrangian function is given by
        \begin{align}\label{eq:proof_Modified_IA_QSMPA_2}
          &\mathcal{L}(\tau^{ (k)}, \{\rho^{(k)}_i\}_{\forall i}) \nonumber \\
          &= y^{(k)} + \tau^{ (k)}\left(\sum_{i=1}^{G}\frac{N_sF}{G}P^{(k)}[i] - P_{\rm tot}\right) \notag\\
            &~~~+ \sum_{i=1}^{G}\rho^{(k)}_i\left(\frac{DB^{(k)}[i]\ln 2}{\Delta f \left(\gamma^{(k)}[i]\alpha_i+\alpha_i-1\right)} - y^{(k)}\right).
        \end{align}
        By applying the stationarity and complementary slackness conditions in the same manner as in Appendix~A, the Lagrange multiplier $\rho^{(k)}_i$ is obtained as
        \begin{align}\label{eq:proof_Modified_IA_QSMPA_3}
            \rho^{(k)}_i = \frac{\tau^{(k)}\sigma^2D\ln2}{\Delta f_0\left\{y^{(k)}\right\}^2}\frac{B^{(k)}[i]}{\alpha_i\lambda^2[i]}.
        \end{align}
         By substituting \eqref{eq:proof_Modified_IA_QSMPA_3} into the normalization condition $\sum_i \rho^{(k)}_i = 1$ yields the closed-form expression for $y^{(k)}$ in \eqref{y_k}.
         As discussed in \eqref{eq:proof_IA_QSMPA_4}, the complementary slackness condition leads to the following expression:        \begin{align}\label{eq:proof_Modified_IA_QSMPA_2_sub}
           y^{(k)} = \frac{DB^{(k)}[i]\ln 2}{\Delta f \left(\gamma^{(k)}[i]\alpha_i+\alpha_i-1\right)}.
        \end{align}
         From this relationship, the corresponding power allocation $P^{(k)}[i]$ can be derived, as shown in \eqref{eq:U_i_1}.

        \item[(ii)] {\bf Case 2} ($1 \leq \gamma^{(k-1)}[i]$): In this case, the Lagrangian function is given by
        \begin{align}\label{eq:proof_Modified_IA_QSMPA_4}
          &\mathcal{L}(\tau^{ (k)}, \{\rho^{(k)}_i\}_{\forall i}) \nonumber \\
          &= y^{(k)} + \tau^{ (k)}\left(\sum_{i=1}^{G}\frac{N_sF}{G}P^{(k)}[i] - P_{\rm tot}\right) \notag\\
            &~~~+ \sum_{i=1}^{G}\rho^{(k)}_i\left(\frac{DB^{(k)}[i]\ln 2}{\Delta f \left(\gamma^{(k)}[i]-2+2\alpha_i\right)} - y^{(k)}\right).
        \end{align}
        By applying the stationarity and complementary slackness conditions in the same manner as in Appendix~A, the Lagrange multiplier $\rho^{(k)}_i$ is obtained as
        \begin{align}\label{eq:proof_Modified_IA_QSMPA_5}
          \rho^{(k)}_i = \frac{\tau^{(k)}\sigma^2D\ln2}{\Delta f_0\left\{y^{(k)}\right\}^2}\frac{B^{(k)}[i]}{\lambda^2[i]}.
        \end{align}
        By substituting \eqref{eq:proof_Modified_IA_QSMPA_4} into the normalization condition $\sum_i \rho^{(k)}_i = 1$ yields the closed-form expression for $y^{(k)}$ in \eqref{y_k}.
         As discussed in \eqref{eq:proof_IA_QSMPA_4}, the complementary slackness condition leads to the following expression:        \begin{align}\label{eq:proof_Modified_IA_QSMPA_5_sub}
           y^{(k)} =\frac{DB^{(k)}[i]\ln 2}{\Delta f \left(\gamma^{(k)}[i]-2+2\alpha_i\right)}.
        \end{align}
         From this relationship, the corresponding power allocation $P^{(k)}[i]$ can be derived, as shown in \eqref{eq:U_i_1}.

    \end{itemize}
    The optimal Lagrange multiplier $\tau^{\star(k)}$ is then determined to satisfy the total power constraint in \eqref{eq:P_k_i_sum}. This completes the proof of Theorem~2.


\bibliographystyle{IEEEtran}
\bibliography{Reference}

\begin{thebibliography}{10}
\providecommand{\url}[1]{#1}
\csname url@samestyle\endcsname
\providecommand{\newblock}{\relax}
\providecommand{\bibinfo}[2]{#2}
\providecommand{\BIBentrySTDinterwordspacing}{\spaceskip=0pt\relax}
\providecommand{\BIBentryALTinterwordstretchfactor}{4}
\providecommand{\BIBentryALTinterwordspacing}{\spaceskip=\fontdimen2\font plus
\BIBentryALTinterwordstretchfactor\fontdimen3\font minus
  \fontdimen4\font\relax}
\providecommand{\BIBforeignlanguage}[2]{{%
\expandafter\ifx\csname l@#1\endcsname\relax
\typeout{** WARNING: IEEEtran.bst: No hyphenation pattern has been}%
\typeout{** loaded for the language `#1'. Using the pattern for}%
\typeout{** the default language instead.}%
\else
\language=\csname l@#1\endcsname
\fi
#2}}
\providecommand{\BIBdecl}{\relax}
\BIBdecl

\bibitem{SC_1}
X.~Luo, H.-H. Chen, and Q.~Guo, ``{Semantic communications: Overview, open
  issues, and future research directions},'' \emph{IEEE Wireless Commun.},
  vol.~29, no.~1, pp. 210--219, Feb. 2022.

\bibitem{Application_2}
D.~Gündüz, Z.~Qin, I.~E. Aguerri, H.~S. Dhillon, Z.~Yang, A.~Yener, K.~K.
  Wong, and C.-B. Chae, ``{Beyond transmitting bits: Context, semantics, and
  task-oriented communications},'' \emph{IEEE J. Sel. Areas Commun.}, vol.~41,
  no.~1, pp. 5--41, Jan. 2023.

\bibitem{Application_3}
C.~Zhang, H.~Zou, S.~Lasaulce, W.~Saad, M.~Kountouris, and M.~Bennis,
  ``{Goal-oriented communications for the IoT and application to data
  compression},'' \emph{IEEE Internet Things Mag.}, vol.~5, no.~4, pp. 58--63,
  Dec. 2022.

\bibitem{image_trans_2}
E.~Erdemir, T.-Y. Tung, P.~L. Dragotti, and D.~Gündüz, ``{Generative joint
  source-channel coding for semantic image transmission},'' \emph{IEEE J. Sel.
  Areas Commun.}, vol.~41, no.~8, pp. 2645--2657, Aug. 2023.

\bibitem{image_trans_3}
D.~B. Kurka and D.~Gündüz, ``{Bandwidth-agile image transmission with deep
  joint source-channel coding},'' \emph{IEEE Trans. Wireless Commun.}, vol.~20,
  no.~12, pp. 8081--8095, Dec. 2021.

\bibitem{DeepSC}
H.~Xie, Z.~Qin, G.~Y. Li, and B.-H. Juang, ``{Deep learning enabled semantic
  communication systems},'' \emph{IEEE Trans. Signal Process.}, vol.~69, pp.
  2663--2675, Apr. 2021.

\bibitem{ReAllo-T}
L.~Yan, Z.~Qin, R.~Zhang, Y.~Li, and G.~Y. Li, ``{Resource allocation for text
  semantic communications},'' \emph{IEEE Wireless Commun. Lett.}, vol.~11,
  no.~7, pp. 1394--1398, Jul. 2022.

\bibitem{DeepSC-S}
Z.~Weng and Z.~Qin, ``{Semantic communication systems for speech
  transmission},'' \emph{IEEE J. Sel. Areas Commun.}, vol.~39, no.~8, pp.
  2434--2444, Aug. 2021.

\bibitem{NECST}
K.~Choi, K.~Tatwawadi, A.~Grover, T.~Weissman, and S.~Ermon, ``{Neural joint
  source-channel coding},'' in \emph{Proc. Int. Conf. Mach. Learn. (ICML)},
  Long Beach, CA, USA, Jun. 2019, pp. 1182--1192.

\bibitem{DSC_Fixed_bit_6}
Q.~Fu, H.~Xie, Z.~Qin, G.~Slabaugh, and X.~Tao, ``{Vector quantized semantic
  communication system},'' \emph{IEEE Commun. Lett.}, vol.~12, no.~6, pp.
  982--986, Jun. 2023.

\bibitem{EC_2}
J.~Huang, K.~Yuan, C.~Huang, and K.~Huang, ``{D$^2$-JSCC: Digital deep joint
  source-channel coding for semantic communications},'' \emph{IEEE J. Sel.
  Areas Commun.}, vol.~43, no.~4, pp. 1246--1261, Apr. 2025.

\bibitem{DSC_Fixed_bit_1}
J.~Park, Y.~Oh, S.~Kim, and Y.-S. Jeon, ``{Joint source-channel coding for
  channel-adaptive digital semantic communications},'' \emph{IEEE Trans. Cogn.
  Commun. Netw.}, vol.~11, no.~1, pp. 75--89, Feb. 2025.

\bibitem{DSC_Fixed_bit_7}
Y.~Oh, J.~Park, J.~Choi, J.~Park, and Y.-S. Jeon, ``{Blind training for
  channel-adaptive digital semantic communications},'' 2025,
  \textit{arXiv:2501.02273v2}.

\bibitem{DSC_Symbol_1}
Y.~Bo, Y.~Duan, S.~Shao, and M.~Tao, ``{Joint coding-modulation for digital
  semantic communications via variational autoencoder},'' \emph{IEEE Trans.
  Commun.}, vol.~72, no.~9, pp. 5626--5640, Sep. 2024.

\bibitem{DSC_Fixed_bit_5}
L.~Guo, W.~Chen, Y.~Sun, and B.~Ai, ``{Digital-SC: Digital semantic
  communication with adaptive network split and learned non-linear
  quantization},'' \emph{IEEE Trans. Cogn. Commun. Netw.}, early access, Dec.
  2024, doi: 10.1109/TCCN.2024.3510586.

\bibitem{FATD}
G.~Zhang, P.~Yang, Y.~Cai, Q.~Hu, and G.~Yu, ``{From analog to digital:
  Multi-order digital joint coding-modulation for semantic communication},''
  \emph{IEEE Trans. Commun.}, early access, Dec. 5, 2024, doi:
  10.1109/TCOMM.2024.3511949.

\bibitem{Digital_JSCC_MIMO_3}
H.~Wu, Y.~Shao, C.~Bian, K.~Mikolajczyk, and D.~Gündüz, ``{Deep joint
  source-channel coding for adaptive image transmission over MIMO channels},''
  \emph{IEEE Trans. Wireless Commun.}, vol.~23, no.~10, pp. 15\,002--15\,017,
  Oct. 2024.

\bibitem{Digital_JSCC_MIMO_5}
M.~Yang, C.~Bian, and H.-S. Kim, ``{OFDM-guided deep joint source channel
  coding for wireless multipath fading channels},'' \emph{IEEE Trans. Cogn.
  Commun. Netw.}, vol.~8, no.~2, pp. 584--599, Jun. 2022.

\bibitem{Digital_JSCC_OFDM}
J.~Park, H.~Kim, J.~Shin, Y.~Oh, and Y.-S. Jeon, ``{End-to-end training and
  adaptive transmission for OFDM-based semantic communication},'' 2025, to be
  appeared in \textit{ICT Express}.

\bibitem{VQA}
H.~Xie, Z.~Qin, and G.~Y. Li, ``{Task-oriented multi-user semantic
  communications for VQA},'' \emph{IEEE Wireless Commun. Lett.}, vol.~11,
  no.~3, pp. 553--557, Mar. 2022.

\bibitem{MDCEI}
J.~Shao, Y.~Mao, and J.~Zhang, ``{Task-oriented communication for multidevice
  cooperative edge inference},'' \emph{IEEE Trans. Wireless Commun.}, vol.~22,
  no.~1, pp. 73--87, Jan. 2023.

\bibitem{MVID}
W.~Xu, Y.~Zhang, F.~Wang, Z.~Qin, C.~Liu, and P.~Zhang, ``{Semantic
  communication for the internet of vehicles: A multiuser cooperative
  approach},'' \emph{IEEE Veh. Technol. Mag.}, vol.~18, no.~1, pp. 100--109,
  Mar. 2023.

\bibitem{TF_IA_1}
L.~Teng, W.~An, C.~Dong, and X.~Xu, ``{sDMCM—semantic digital modulation
  constellation mapping scheme for semantic communication},'' \emph{IEEE
  Internet Things J.}, early access, Feb. 2025, doi: 10.1109/JIOT.2025.3545667.

\bibitem{TF_IA_2}
H.~Gao, G.~Yu, Y.~He, and Y.~Liu, ``{Semantic feature scheduling and rate
  control in multi-modal distributed network},'' \emph{IEEE Trans. Wireless
  Commun.}, vol.~23, no.~12, pp. 19\,199--19\,214, Dec. 2024.

\bibitem{TF_IA_3}
K.~Zhou, G.~Zhang, Y.~Cai, Q.~Hu, G.~Yu, and A.~L. Swindlehurst, ``{Feature
  allocation for semantic communication with space-time importance
  awareness},'' \emph{IEEE Trans. Wireless Commun.}, early access, May 2025,
  doi: 10.1109/TWC.2025.3569320.

\bibitem{TF_IA_4}
J.~Im, N.~Kwon, T.~Park, J.~Woo, J.~Lee, and Y.~Kim, ``{Attention-aware
  semantic communications for collaborative inference},'' \emph{IEEE Internet
  Things J.}, vol.~11, no.~22, pp. 37\,008--37\,020, Nov. 2024.

\bibitem{TF_IA_5}
J.~Park, Y.~Oh, Y.~Kim, and Y.-S. Jeon, ``{Vision transformer-based semantic
  communications with importance-aware quantization},'' 2024,
  \textit{arXiv:2412.06038}.

\bibitem{BCD}
L.~Peng and R.~Vidal, ``{Block coordinate descent on smooth manifolds:
  Convergence theory and twenty-one examples},'' 2023,
  \textit{arXiv:2305.14744}.

\bibitem{MVP_N}
R.~Wang, T.~S. Kim, J.-S. Kim, and H.-J. Lee, ``{Towards real-world multi-view
  object classification: dataset, benchmark, and analysis},'' \emph{IEEE Trans.
  Circuits Syst. Video Technol.}, vol.~34, no.~7, pp. 5653--5664, Jul. 2024.

\bibitem{Block_fading}
A.~Goldsmith, \emph{{Wireless communications}}.\hskip 1em plus 0.5em minus
  0.4em\relax Cambridge, U.K.: Cambridge Univ. Press, 2005.

\bibitem{Water_filling_3}
X.~Ling, B.~Wu, P.-H. Ho, F.~Luo, and L.~Pan, ``{Fast water-filling for agile
  power allocation in multi-channel wireless communications},'' \emph{IEEE
  Commun. Lett.}, vol.~16, no.~8, pp. 1212--1215, Aug. 2012.

\bibitem{Water_filling_2}
F.~Gao, T.~Cui, and A.~Nallanathan, ``{Optimal training design for channel
  estimation in decode-and-forward relay networks with individual and total
  power constraints},'' \emph{IEEE Trans. Signal Process.}, vol.~56, no.~12,
  pp. 5937--5949, Dec. 2008.

\bibitem{Finite_block_cap}
Z.~Ma, M.~Xiao, Y.~Xiao, Z.~Pang, H.~V. Poor, and B.~Vucetic,
  ``{High-reliability and low-latency wireless communication for internet of
  things: Challenges, fundamentals, and enabling technologies},'' \emph{IEEE
  Internet Things J.}, vol.~6, no.~5, pp. 7946--7970, Oct. 2019.

\bibitem{D_model}
H.~Touvron, M.~Cord, M.~Douze, F.~Massa, A.~Sablayrolles, and H.~Jégou,
  ``{Training data-efficient image transformers \& distillation through
  attention},'' in \emph{Proc. Int. Conf. Mach. Learn. (ICML)}, Jul. 2021, pp.
  10\,347--10\,357.

\end{thebibliography}

\end{document}